\documentclass[11pt]{article}
\usepackage[margin=2.5cm,top=2cm,bottom=2.5cm,centering]{geometry}
\usepackage[utf8]{inputenc}
\usepackage[english]{babel}
\usepackage{graphicx}
\usepackage{amsmath}
\usepackage{amsfonts}
\usepackage{amssymb}
\usepackage{amsthm}
\usepackage{mathrsfs}
\usepackage{enumerate}
\usepackage[hidelinks]{hyperref}
\usepackage{tikz}
\usepackage[hypcap]{caption}
\usepackage{bm}
\usepackage{cancel}
\usepackage{cite}
\usepackage{lscape}
\usetikzlibrary{calc}
\usetikzlibrary{decorations.pathmorphing}
\usetikzlibrary{arrows,shapes,positioning}
\usetikzlibrary{decorations.markings}
\usetikzlibrary{patterns}
\usetikzlibrary{lindenmayersystems}

\hypersetup{
colorlinks=true,
citecolor=red,
filecolor=black,
linkcolor=blue,
urlcolor=blue,
pdfauthor={Adrien Poncelet and Philippe Ruelle},
pdftitle={Sandpile probabilities on triangular and hexagonal lattices},
pdfstartview={XYZ null null 1.50},
bookmarksopen=true
}

\numberwithin{equation}{section}

\newcommand*\xbar[1]{%
  \hbox{%
    \vbox{%
      \hrule height 0.5pt 
      \kern0.3ex
      \hbox{%
        \kern-0.1em
        \ensuremath{#1}%
        \kern-0.0em
      }%
    }%
  }%
} 

\newcommand{\C}{\mathbb{C}}
\newcommand{\Z}{\mathbb{Z}}

\newcommand{\G}{\mathcal{G}}
\newcommand{\N}{\mathcal{N}}

\newcommand{\Gr}{\mathbf{G}}
\newcommand{\Zr}{\mathbf{Z}}
\newcommand{\bs}{\backslash}
\newcommand{\diff}{\text{d}}

\renewcommand{\i}{\text{i}}
\renewcommand{\l}{\ell}
\renewcommand{\P}{\mathbb{P}}

\renewcommand{\ge}{\geqslant}
\renewcommand{\le}{\leqslant}

\title{\bf Sandpile probabilities on triangular\\and hexagonal lattices}
\author{\normalsize \textsc{Adrien Poncelet}, \textsc{Philippe Ruelle}\medskip\\
{\normalsize
\begin{minipage}{0.95\textwidth}
\begin{center}
\textit{Universit\'e catholique de Louvain\\Institut de recherche en math\'ematique et physique\\Chemin du Cyclotron 2, 1348 Louvain-la-Neuve, Belgium}\\
\medskip
\href{mailto:adrien.poncelet@uclouvain.be}{\normalsize\texttt{adrien.poncelet@uclouvain.be}}, \href{mailto:philippe.ruelle@uclouvain.be}{\normalsize\texttt{philippe.ruelle@uclouvain.be}}
\end{center}
\end{minipage}}
}
\date{}

\begin{document}
\maketitle

\begin{abstract}
We consider the Abelian sandpile model on triangular and hexagonal lattices. We compute several height probabilities on the full plane and on half-planes, and discuss some properties of the universality of the model. 

\end{abstract}


\section{Introduction}
\label{sec1}

The Abelian sandpile model, introduced by Bak, Tang and Wiesenfeld \cite{BTW87}, is one of the most studied complex system exhibiting \emph{self-organized criticality}, meaning it naturally evolves toward and maintains itself in a critical state. Over the years, many analytical and numerical results have been obtained; the two-dimensional isotropic model being particularly amenable to computations due to its Abelian property \cite{Dha90}. Among them are single-site height probabilities on the full lattice or half-lattice \cite{Pri94,JPR06}, various joint probabilities of heights at isolated sites or between a certain type of clusters \cite{MD91,MR01,Jen05a,PGPR08,PGPR10}, the effects of boundary conditions and boundary probabilities \cite{BIP93,Iva94,Jen04,Jen05b,PR05a,Rue07}, and avalanche and toppling wave distributions \cite{IKP94a,IKP94b,PKI96,KP98}. The model has also been studied, in the scaling limit, in a field-theoretical framework. Namely, it has been shown to be described by a logarithmic conformal field theory with central charge $c=-2$ \cite{MD92,MR01,MRR05,PR05b,JPR06} (for a recent review, see also \cite{Rue13}).

Most of the statistical results have been obtained for a sandpile model defined on the simplest two-dimensional regular lattice, namely the square grid. However, other regular graphs may also be considered, such as the triangular and hexagonal (or honeycomb) lattices, to check the universality of the Abelian sandpile model. In particular, height probabilities and critical exponents have been investigated on such lattices in \cite{PP96,LH02,HL03} using renormalization group transformations and numerical simulations. More recently, multisite height-one probabilities have been evaluated exactly on the hexagonal lattice \cite{ADMR10}. All the results so far point toward a common set of critical exponents and scaling behavior for all lattices.

In this paper, we aim to provide further verifications of the universality of the Abelian sandpile model. To do so, we compute one-site height probabilities on the triangular and hexagonal graphs, for the full infinite lattice or half-lattices. We also give explicit expressions for two-site probabilities on the boundary of two types of hexagonal half-lattices. 

The article is organized as follows. In Section \ref{sec2}, we recall the definition of the Abelian sandpile model on a generic graph. In Section \ref{sec3}, we discuss the relation between recurrent states of the model and uniform spanning forests with marked vertices called \emph{nodes}. We give an overview of the techniques required to compute probabilities in the latter model, based on graphs equipped with a \emph{connection} \cite{For93,Ken11,KW15}. Using these results, we compute several explicit sandpile probabilities on the triangular and hexagonal lattices in Sections \ref{sec4} and \ref{sec5}. We compare our results with their analogues on the square lattice in Section \ref{sec6}.


\section{The Abelian sandpile model}
\label{sec2}
In this section we recall the definition of the Abelian sandpile model \cite{BTW87} on a generic undirected connected graph $\G$. Let us denote by $\G^*$ the extended graph obtained from $\G$ by adding a vertex $s$ together with edges between $s$ and a collection of selected vertices $\mathcal{D}$ of $\G$. The vertex $s$ is called a \emph{sink} or \emph{root}, and its neighbors on $\G^*$ (i.e. the vertices in $\mathcal{D}$) are called \emph{dissipative} or \emph{open} sites ($\mathcal{D}$ is also said to be \emph{wired} to the sink). By contrast, the vertices in $\G\backslash\mathcal{D}$ are said to be \emph{closed}. In all computations presented in this paper, we shall assume $\G$ to be planar, such that $\mathcal{D}$ is a subset of the vertices adjacent to the outer face of the graph.

With each vertex or \emph{site} $i$ of $\G$ we associate an integer $h_i\in\mathbb{N}^*$, which can be thought of as the number of sand grains piled up at $i$. We call the collection $\mathcal{C}=\left\{h_i\right\}_{i\in\G}$ a \emph{sandpile configuration} on $\G$. $\mathcal{C}$ is said to be \emph{stable} if $h_i\le\deg_{\G^*}(i)$ for every site $i\in\G$, where $\deg_{\G^*}(i)$ is the degree of $i$ in $\G^*$. The \emph{toppling matrix} $\Delta=\Delta_{\G}$ of the Abelian sandpile model is defined as follows, for any $i,j\in\G$:
\begin{equation}
\Delta_{i,j}=\begin{cases}
\deg_{\G^*}(i)&\quad\textrm{if $i=j$,}\\
-1&\quad\textrm{if $i$ and $j$ are neighbors on $\G$,}\\
0&\quad\textrm{otherwise.}
\end{cases}
\label{top_mat}
\end{equation}
Hence, the toppling matrix is the discrete graph Laplacian with Dirichlet boundary conditions at $\{s\}$ (that is, the standard graph Laplacian of $\G^*$ from which we remove the row and column indexed by $s$).

The sandpile model evolves in time through a two-step discrete algorithm, mapping a stable configuration $\mathcal{C}_t$ at time $t$ onto another stable configuration $\mathcal{C}_{t+1}$ at time $t{+}1$. First a grain of sand is added on a random site $i$, increasing its height $h_i^{\textrm{old}}$ by $1$: $h_i^{\textrm{new}}=h_i^{\textrm{old}}+1$. If $h_i^{\textrm{new}}\le\deg_{\G^*}(i)$, the new configuration is stable and defines $\mathcal{C}_{t+1}$. If rather $h_i^{\textrm{new}}>\deg_{\G^*}(i)$, $i$ \emph{topples}, that is, the height at every site $j$ of $\G$ is shifted by $-\Delta_{i,j}$ (in particular $h_i^{\textrm{new}}\to h_i^{\textrm{new}}-\deg_{\G^*}(i)$). This relaxation procedure is repeated in case some of the shifted heights $h_k$ exceed $\deg_{\G^*}(k)$, until all sites of $\G$ are stable. The resulting sandpile configuration defines $\mathcal{C}_{t+1}$.

The sink $s$ plays a particular role, as it never topples (therefore, its height grows unboundedly over time). Its presence is required to ensure that only a finite number of topplings occur at each time step \cite{Dha90}. The model is said to be \emph{Abelian} because the stable configuration obtained through relaxation does not depend on the order in which unstable sites topple.

The dynamics presented above defines a discrete Markov chain on the space of stable configurations. It has a unique invariant measure $\P_{\G}^*$ given by
\begin{equation}
\P_{\G}^*(\mathcal{C})=\begin{cases}
\frac{1}{|\mathcal{R}|}&\quad\textrm{if $\mathcal{C}$ is recurrent,}\\
0&\quad\textrm{if $\mathcal{C}$ is transient,}
\end{cases}
\end{equation}
where $\mathcal{R}$ is the subset of \emph{recurrent} configurations on $\G$, with $|\mathcal{R}|=\det\Delta$ \cite{Dha90}. While $\P_{\G}^*$ is simple since it is uniform on $\mathcal{R}$, it has a complicated support. Indeed, a stable configuration $\mathcal{C}$ is recurrent if and only if it contains no subconfigurations of a certain type, said to be \emph{forbidden} \cite{Dha90}. Assessing whether a configuration is recurrent or transient therefore requires that the whole lattice be scanned.

An alternative, more useful, characterization of recurrent sandpile configurations has been provided by the so-called \emph{burning algorithm} \cite{MD92}, which yields a bijection between such configurations on $\G$ and uniform spanning trees on $\G^*$. Most of the sandpile results have been obtained through this correspondence, in particular single-site and multisite height probabilities. Indeed, this algorithm allows one to trade a globally constrained model in terms of local degrees of freedom for an unconstrained one in terms of nonlocal degrees of freedom; the latter of which proving to be analytically more tractable. The calculations with standard graph-theoretical methods have however remained long and cumbersome. Recently, Kenyon and Wilson have devised new techniques using graphs with connections \cite{KW15}, which greatly simplify spanning tree computations. In the next section, we recall the explicit equation relating height probabilities to certain classes of spanning trees, and give an overview of the methods developed in \cite{KW15}. We refer the reader to our previous work \cite{PR17} for a detailed exposition of these techniques applied specifically to the sandpile model on the square lattice.


\section{Spanning tree probabilities}
\label{sec3}
The authors of \cite{MD92} defined a burning algorithm to establish a one-to-one correspondence between recurrent sandpile configurations on a graph $\G$ and rooted spanning trees on $\G^*=\G\cup\{s\}$, that is, oriented spanning trees in which all edges are directed toward the root $s$. A modified version of this algorithm was used in \cite{Pri94} to express one-site sandpile probabilities $\P^{\G}_a(i)\equiv\P^{\G}(h_i{=}a)$ in terms of fractions of spanning trees with fixed topological properties:
\begin{equation}
\P^{\G}_a(i) = \sum_{q=0}^{a-1}\frac{X^{\G}_q(i)}{\deg_{\G^*}(i)-q} = \P^{\G}_{a-1}(i) + \frac{X^{\G}_{a-1}(i)}{\deg_{\G^*}(i)+1-a}, \qquad (\P^{\G}_0(i)\equiv0),
\label{P_X}
\end{equation}
where $X^{\G}_q(i)$ denotes the fraction of rooted spanning trees on $\G^*$ such that $i$ has $q$ \emph{predecessors} among its nearest neighbors ($j$ is a predecessor of $i$ on a rooted spanning tree if the unique path from $j$ to the root $s$ goes through $i$). A similar decomposition of spanning trees into subclasses can be utilized for multisite probabilities \cite{Jen05b,PR05a,PR17}.

Among the fractions of spanning trees $X^{\G}_q(i)$, the case $q=0$ is the easiest to compute (see \cite{MD91} for the square lattice, and \cite{ADMR10} for the hexagonal lattice). Trees contributing to $X^{\G}_0(i)$ are such that $i$ is a \emph{leaf}, as it has no predecessor among its nearest neighbors. Such spanning trees on $\G^*$ are in $\deg_{\G^*}(i)$-to-one correspondence with unconstrained spanning trees on a modified graph $\xbar{\G}^*$, obtained by cutting all but one adjacent edge to $i$. We denote these removed edges by $\{i,j_{\l}\}$, with $1\le\l\le\deg_{\G^*}(i){-}1$. The factor $\deg_{\G^*}(i)$ in the correspondence comes from the fact that a leaf at $i$ can be connected to any of its neighbors on $\G^*$, but to only one on $\xbar{\G}^*$. Therefore
\begin{equation}
X^{\G}_0(i)\det\Delta_{\G}=\deg_{\G^*}(i)\times\det\Delta_{\scriptsize\xbar{\G}}.
\end{equation}
Note that the difference between the two Laplacians, $B=\Delta_{\scriptsize\xbar{\G}}-\Delta_{\G}$, is nonzero on a finite submatrix only: $B_{i,i}=-\deg_{\G^*}(i){+}1$, $B_{j_{\l},j_{\l}}=-1$, $B_{i,j_{\l}}=B_{j_{\l},i}=1$ and $B_{m,n}=0$ otherwise. Its rank is equal to $r=\deg_{\G^*}(i){-}1$. Hence, the fraction of interest reads:
\begin{equation}
X^{\G}_0(i)=\deg_{\G^*}(i)\times\det(\mathbb{I}+GB),
\label{X0_comp}
\end{equation}
where $G\equiv(\Delta_{\G})^{-1}$ is the Green function of the graph $\G^*$ with Dirichlet boundary conditions at $\{s\}$. Since the rank of $B$ is finite, one sees that the computation of $X^{\G}_0(i)$ consists in evaluating the determinant of an $r\times r$ matrix, even if the graph $\G$ is infinite. Finding the explicit value of $X^{\G}_0(i)$ on regular graphs is therefore straightforward, as the Green functions for suchs graphs are well known (see e.g. \cite{Cse00}, and the appendix of this paper for the triangular lattice).

Fractions $X^{\G}_{q>0}(i)$ are harder to evaluate because the predecessor property imposes nonlocal constraints on spanning trees. A closed form for one-point probabilities $\P_a(i)$ on the square lattice was only obtained \cite{PPR11,CS12,KW15} twenty years after they were first written using an integral representation \cite{Pri94}. In particular the authors of \cite{KW15} developed a combinatorial method based on the so-called \emph{line bundle Laplacian} introduced in \cite{Ken11} (see also \cite{For93}), which we have used to obtain the results presented in this article. We shall give an overview of their method, and refer the reader to the original paper for more details, or to \cite{PR17} for its specific application to the Abelian sandpile model.

The main tool of \cite{KW15}, the line bundle Laplacian $\mathbf{\Delta}=\mathbf{\Delta}_{\G}$, is an extension of the usual graph Laplacian $\Delta$. It is defined as follows, for Dirichlet boundary conditions at $\{s\}$ as in Section~\ref{sec2},
\begin{equation}
\mathbf{\Delta}_{i,j}=\begin{cases}
\deg_{\G^*}(i)&\quad\textrm{if $i=j$,}\\
-\phi_{j,i}&\quad\textrm{if $i$ and $j$ are neighbors on $\G$,}\\
0&\quad\textrm{otherwise,}
\end{cases}
\label{lbl}
\end{equation}
for any vertices $i,j\in\G$ (thus excluding the sink $s$). The $\phi_{i,j}$'s are nonzero complex numbers that satisfy the property $\phi_{j,i}=\phi_{i,j}^{-1}$. The quantity $\phi_{i,j}$ is called the \emph{parallel transport from $i$ to $j$}. If all parallel transports are trivial, $\mathbf{\Delta}=\Delta$. This generalized\footnote{It should be noted that the definition \eqref{lbl} differs from that of a weighted graph Laplacian, since the $\phi_{i,j}$'s do not contribute to the diagonal of $\mathbf{\Delta}$.} Laplacian allows one to compute specific spanning forest events described hereafter.

\begin{figure}[t]
\centering
\newcommand*\rows{6}
\begin{tikzpicture}
\foreach \row in {0,1,...,\rows} {
\draw[help lines,ultra thin] ($\row*(0.5,{0.5*sqrt(3)})$)--($(\rows,0)+\row*(-0.5,{0.5*sqrt(3)})$);
\draw[help lines,ultra thin] ($\row*(1,0)$)--($(\rows/2,{\rows/2*sqrt(3)})+\row*(0.5,{-0.5*sqrt(3)})$);
\draw[help lines,ultra thin] ($\row*(1,0)$)--($(0,0)+\row*(0.5,{0.5*sqrt(3)})$);}
\foreach \x in {0,...,6}{
\draw[help lines,ultra thin,shift={(6,0)}] (120:\x)--++(0:0.25);
\draw[help lines,ultra thin,shift={(6,0)}] (120:\x)--++(60:0.25);
\draw[help lines,ultra thin,shift={(60:\x)}] (0,0)--++(120:0.25);
\draw[help lines,ultra thin,shift={(60:\x)}] (0,0)--++(180:0.25);
\draw[help lines,ultra thin,shift={(\x,0)}] (0,0)--++(240:0.25);
\draw[help lines,ultra thin,shift={(\x,0)}] (0,0)--++(300:0.25);}
\draw[very thick,blue,shift={(-0.375,{-sqrt(3)/8})}] (0,0)--(0:6.75)--(60:6.75)--(0,0);
\filldraw[blue] ($(4.5,{3*sqrt(3)/2})+(30:0.25)$) circle (0.1cm) node[above right] {\large $4$};
\filldraw[blue!15!white] ([xshift=2cm]60:2)--++(60:1)--++(180:1)--++(300:1);
\draw[very thick,blue] ([xshift=2cm]60:2)--++(60:1)--++(180:1)--++(300:1);
\filldraw[blue] ([xshift=2cm]60:2) circle (0.1cm) node[below right] {\large $1$};
\filldraw[blue] ([xshift=2cm]60:3) circle (0.1cm) node[above right] {\large $2$};
\filldraw[blue] ([xshift=1cm]60:3) circle (0.1cm) node[above left] {\large $3$};
\draw[very thick,green!50!black] (3,{4/sqrt(3)})--({4/3},{4/sqrt(3)})--++(-0.25,0);
\draw[very thick,->,green!50!black] ([shift={(1.5,{3*sqrt(3)/2})}]300:0.125)--++(300:0.5);
\draw[very thick,red] (3,{4/sqrt(3)})--({5/3},0)--++(240:0.25);
\filldraw (3,{4/sqrt(3)}) circle (0.1cm);
\draw[very thick,->,red] (1.9,{sqrt(3)/2})--(2.4,{sqrt(3)/2});
\end{tikzpicture}
\caption{Annular-one surface graph on a triangular grid with wired boundary (the triangular box is identified as the sink). Three nodes lie around a single face $f$ (the grey face in the center), while the fourth one is taken to be the sink. Two admissible zippers from $f$ to the outer face are represented as straight lines for simplicity, instead of paths on the dual hexagonal graph. Their arrow indicates the orientation of the parallel transport $z$ on the zipper edges.}
\label{an1_ex}
\end{figure}
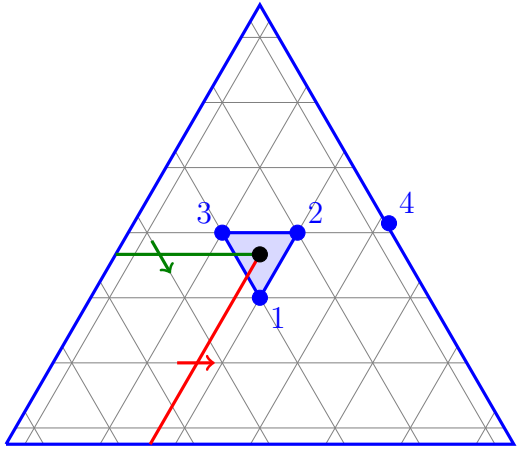

Let $\G^*$ be a planar graph with a marked face $f$. We select a subset $\N$ of the vertices lying along the boundary of $f$, together with the sink $s$, and call them \emph{nodes}. We label them from 1 to $n{-}1$ in counterclockwise order, with $n=s$. In \cite{KW15} $\G^*$, together with $\N$, is called an \emph{annular-one surface graph}. Let $\mathbf{\Delta}$ be the line bundle Laplacian on $\G^*$ with trivial parallel transports everywhere except on the directed edges $(k,\l)$ crossed by a \emph{zipper}, that is, a path on the dual graph from $f$ to the outer face: $\phi_{k,\l}=\phi_{\l,k}^{-1}=z\in\C^*$. It follows that the Laplacian $\mathbf{\Delta}(z)$ and its inverse $\Gr(z)=[\mathbf{\Delta}(z)]^{-1}$ depend on this parameter $z$, with $\mathbf{\Delta}(1)=\Delta$ and $\Gr(1)=G$. The function $\Gr(z)$ is called the \emph{line bundle Green function} of the graph $\G^*$. Up to relabeling, we can always assume that the edge adjacent to $f$ that the zipper crosses lies between nodes $1$ and $n{-}1$, as depicted in Fig.~\ref{an1_ex}.

\begin{figure}[t]
\centering
\newcommand*\rows{6}
\begin{tikzpicture}
\foreach \row in {0,1,...,\rows}{
\draw[help lines,ultra thin] ($\row*(0.5,{0.5*sqrt(3)})$)--($(\rows,0)+\row*(-0.5,{0.5*sqrt(3)})$);
\draw[help lines,ultra thin] ($\row*(1,0)$)--($(\rows/2,{\rows/2*sqrt(3)})+\row*(0.5,{-0.5*sqrt(3)})$);
\draw[help lines,ultra thin] ($\row*(1,0)$)--($(0,0)+\row*(0.5,{0.5*sqrt(3)})$);}
\foreach \x in {0,...,6}{
\draw[help lines,ultra thin,shift={(6,0)}] (120:\x)--++(0:0.25);
\draw[help lines,ultra thin,shift={(6,0)}] (120:\x)--++(60:0.25);}
\draw[ultra thick,blue] (60:1)--(0,0)--++(0:1)--++(60:1)--++(120:1);
\draw[ultra thick,orange] (60:6)--++(0:0.25);
\draw[ultra thick,orange] ([xshift=2cm]60:2)--++(120:1)--++(60:1)--++(120:1)--++(0:1.25);
\draw[ultra thick,orange] ([xshift=2cm]60:3)--++(60:1.25);
\draw[ultra thick,orange] ([xshift=4cm]60:2)--++(240:1)--++(0:1)--++(60:0.25);
\draw[ultra thick,orange] (0:6.25)--(60:6.25);
\draw[ultra thick,green!50!black] (60:4)--++(240:1)--++(300:2)--++(240:1);
\draw[ultra thick,green!50!black] ([xshift=2cm]60:1)--++(300:1)--++(0:3);
\draw[ultra thick,green!50!black] (0:4)--++(120:1)--++(60:2);
\filldraw[blue] (60:1) circle (0.1cm) node[above left] {\large $1$};
\filldraw[blue] ([xshift=1cm]60:1) circle (0.1cm) node[below right] {\large $4$};
\filldraw[orange] ([xshift=1cm]60:3) circle (0.1cm) node[below left] {\large $2$};
\filldraw[orange,shift={(3,0)}] (60:3.25) circle (0.1cm) node[above right] {\large $6$};
\filldraw[green!50!black] (0:3) circle (0.1cm) node[below left] {\large $5$};
\filldraw[green!50!black] ([xshift=3cm]60:1) circle (0.1cm) node[above left] {\large $3$};
\end{tikzpicture}
\caption{Spanning forest on a triangular grid with a wired boundary on its right side, consisting in three components and containing six nodes (the sixth one being the sink $s$). The nodes are distributed in the trees according to the partition $\sigma=14|35|26$.}
\label{node_partition}
\end{figure}
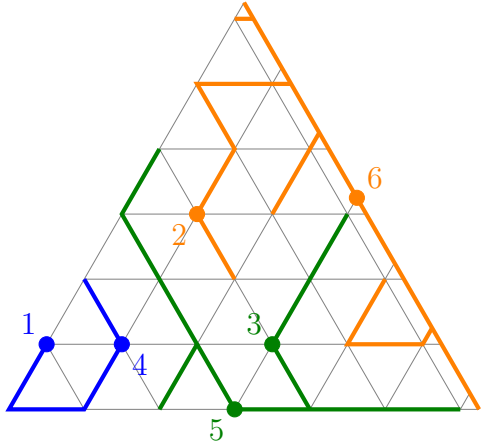

Let $\sigma$ be a partition of the nodes $\N$ such that $n$ (the sink) is not in a singleton. We denote by $Z[\sigma]$ the number of spanning forests on $\G^*$ such that the nodes $\{1,\ldots,n\}$ lie on different trees according to $\sigma$ (see for instance Fig.~\ref{node_partition}); so $Z \equiv Z[12\cdots n]$ is the total number of spanning trees on $\G^*$. As we shall explain below, the fractions of spanning trees $X^{\G}_q(i)$, and hence the one-site sandpile probabilities, can be expressed in terms of the $Z[\sigma]$'s. For partial pairings, that is, partitions of the form $\sigma=r_1 s_1|\cdots|r_k s_k|t_1|\cdots|t_{\l}$, with $\N=R\cup S\cup T$ and $R,S,T$ disjoint, the all-minors matrix-tree theorem \cite{Che76} states that
\begin{equation}
Z\det G_{R\cup T}^{S\cup T}=\sum_{\rho\in\text{S}_k}\varepsilon(\rho)\,Z[r_1 s_{\rho(1)}|\cdots|r_k s_{\rho(k)}|t_1|\cdots|t_{\l}],
\label{AMMT}
\end{equation}
where the sum is over all permutations of the symmetric group on $k$ objects, and $\varepsilon(\rho)$ is the signature of $\rho$. The notation $G_{R\cup T}^{S\cup T}$ refers to the restriction of the Green matrix to the rows (resp. columns) indexed by the nodes in $R\cup T$ (resp. $S\cup T$), with the condition $G_{i,n}=G_{n,i}=1$ for any node $1\le i\le n{-}1$.

For annular-one surface graphs, there are $\frac{1}{2}\left({2k\atop k}\right)$ possible planar pairings of $2k$ nodes (assuming the set $T$ of isolated nodes is fixed, and does not contain the sink $n$), and as many ways to divide $\N\bs T$ into two subsets of order $k$. However, the linear system for the $Z[\sigma]$'s provided by Eq.~\eqref{AMMT} is not invertible. A solution to this problem has been presented in \cite{KW15}. It relies on a generalization of the all-minors matrix-tree theorem for graphs with a nontrivial connection, which relates minors of the line bundle Green function to partition functions of combinatorial objects called \emph{cycle-rooted groves}. These are spanning subgraphs in which each component is either a tree with at least one node, or a cycle-rooted tree (i.e. a unicycle) containing no nodes. The partition function of all cycle-rooted groves in which the nodes are distributed in the tree components according to the partition $\sigma$ is denoted by $\Zr[\sigma]$. The weight of a grove is given by the product of the weights of its components. The weight of a spanning tree is defined as the product of the weights of its edges, while that of a cycle-rooted tree includes an extra factor that vanishes for a trivial connection. For such a connection, a cycle-rooted grove is therefore simply a spanning forest, and $\Zr[\sigma]=Z[\sigma]$.

The key property of this generalized version of Eq.~\eqref{AMMT} is that the linear system it provides becomes invertible on an annular-one surface graph equipped with a nontrivial connection. Hence, the partition function $\Zr[\sigma]$ for any partial pairing can be written as a linear combination of minors of the line bundle Green function $\Gr$. For a connection supported on a zipper with parallel transport $z\in\C^*$, the number of spanning forests $Z[\sigma]$ can then be obtained by taking the limit $z\to 1$ of $\Zr[\sigma]$. The main result of \cite{KW15} is the following:
\begin{equation}
\frac{Z[\sigma]}{Z}\text{ is a function of $G,G'$ evaluated at the nodes $\N\bs\{n\}$,}
\label{grove_thm}
\end{equation}
where $G'=\diff\Gr/\diff z|_{z=1}$ is called the \emph{Green function derivative} (a complicated but explicit combinatorial expression of $Z[\sigma]/Z$ for any partial pairing $\sigma$ has been given in \cite{PW17}). This function can be expressed in terms of the standard Green function through the relation
\begin{equation}
G'_{i,j}=\sum_{(k,\l):\,\phi_{k,\ell}=z}\big(G_{i,\l}\,G_{k,j}-G_{i,k}\,G_{\l,j}\big).
\label{gp_def}
\end{equation}
More generally, the number $Z[\pi]$ of spanning forests whose nodes are distributed in trees according to the generic partition $\pi$ can be computed as a linear combination of $Z[\sigma]$'s (provided that $n$ is not in a singleton), where $\sigma$ are partial pairings \cite{KW15}. We shall give an example of such a reduction in the next section.

Before going any further, let us make two observations. First, a  natural question is, why bother to introduce a connection on the graph and then take its trivial limit? Although the operation seems trivial, it is not, as the ratios $Z[\sigma]/Z$ are functions of the Green function derivative $G'$, which explicitly depends on the location of the zipper edges (see Eq.~\eqref{gp_def}). This dependence remains manifest even after taking the limit $z\to 1$, see \eqref{ztoone} for an illustration. Second, requiring the zipper to start between the nodes $1$ and $n{-}1$ on the boundary of $f$ does not determine it uniquely. Indeed, many choices are admissible (see for instance Fig.~\ref{an1_ex}), and lead to seemingly distinct expressions for the Green function derivative \eqref{gp_def}. It turns out that the functions $G'_{u,v}$ associated with different admissible zippers are closely related to one another: either they are identical, or they differ by $\pm G_{u,v}$ \cite{KW15,PR17}. In any case, they yield the same values for the ratios $Z[\sigma]/Z$.

As the standard Green functions of regular graphs are well known, most of the work consists in evaluating the Green function derivative. We shall discuss two methods for calculating $G'_{u,v}$, depending on whether its arguments $u,v$ are close to or far away from the marked face $f$. In the former case, a clever use \cite{KW15} of the transformation properties of the derivative under translations and deformations of the zipper allows one to find the value $G'_{u,v}$ without actually computing the (possibly infinite) sum appearing in Eq.~\eqref{gp_def}. If rather $u$ and/or $v$ are separated by a large distance $r$ from $f$, we shall give an asymptotic expression for $G'_{u,v}$ in terms of inverse powers of $r$. Since the discussion for both cases is rather technical, we shall reserve it for the appendix.

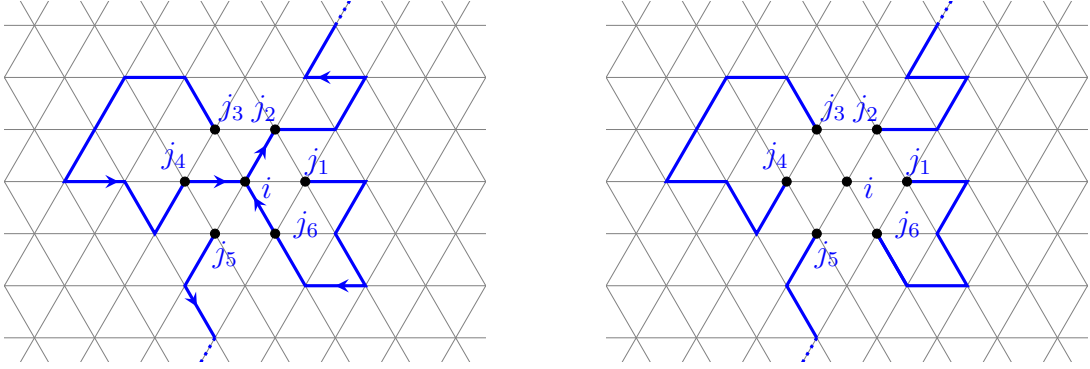
\begin{figure}[t]
\centering
\begin{tikzpicture}[scale=0.8]
\tikzstyle arrowstyle=[scale=1]
\tikzstyle directed=[postaction={decorate,decoration={markings,mark=at position 0.7 with {\arrow[arrowstyle]{stealth}}}}]

\begin{scope}
\clip (-4,-3) rectangle (4,3);
\foreach \x in {-5,-4,...,5}{
\draw[help lines] (-5,{\x*sqrt(3)/2})--(5,{\x*sqrt(3)/2});
\draw[xshift=\x cm,help lines] (60:6)--(240:6);
\draw[xshift=\x cm,help lines] (120:6)--(300:6);
}
\draw[very thick,directed,blue] (120:1)--++(120:1)--++(180:1)--++(240:1)--++(240:1)--++(0:1)--++(300:1)--++(60:1);
\draw[very thick,directed,blue] (180:1)--(0,0);
\draw[very thick,directed,blue] (0:1)--++(0:1)--++(240:1)--++(300:1)--++(180:1)--++(120:1);
\draw[very thick,directed,blue] (300:1)--(0,0);
\draw[very thick,directed,blue] (0,0)--(60:1);
\draw[very thick,directed,blue] (60:1)--++(0:1)--++(60:1)--++(180:1)--++(60:1);
\draw[very thick,dotted,blue] (60:3)--(60:4);
\draw[very thick,directed,blue] (240:1)--++(240:1)--++(300:1);
\draw[very thick,dotted,blue,shift={(1,0)}] (240:3)--(240:4);
\filldraw (0,0) circle (0.075cm);
\foreach \x in {0,1,...,5}{
\filldraw ({\x*60}:1) circle (0.075cm);
}
\node[blue] at (345:0.375) {\large $i$};
\node[blue] at (15:1.25) {\large $j_1$};
\node[blue] at (75:1.25) {\large $j_2$};
\node[blue] at (100:1.25) {\large $j_3$};
\node[blue] at (160:1.25) {\large $j_4$};
\node[blue] at (255:1.25) {\large $j_5$};
\node[blue] at (325:1.25) {\large $j_6$};
\end{scope}

\begin{scope}[xshift=10cm]
\clip (-4,-3) rectangle (4,3);
\foreach \x in {-5,-4,...,5}{
\draw[help lines] (-5,{\x*sqrt(3)/2})--(5,{\x*sqrt(3)/2});
\draw[xshift=\x cm,help lines] (60:6)--(240:6);
\draw[xshift=\x cm,help lines] (120:6)--(300:6);
}
\draw[very thick,blue] (120:1)--++(120:1)--++(180:1)--++(240:1)--++(240:1)--++(0:1)--++(300:1)--++(60:1);
\draw[very thick,blue] (0:1)--++(0:1)--++(240:1)--++(300:1)--++(180:1)--++(120:1);
\draw[very thick,blue] (300:1)--++(300:1);
\draw[very thick,blue] (60:1)--++(0:1)--++(60:1)--++(180:1)--++(60:1);
\draw[very thick,dotted,blue] (60:3)--(60:4);
\draw[very thick,blue] (240:1)--++(240:1)--++(300:1);
\draw[very thick,dotted,blue,shift={(1,0)}] (240:3)--(240:4);
\filldraw (0,0) circle (0.075cm);
\foreach \x in {0,1,...,5}{
\filldraw ({\x*60}:1) circle (0.075cm);
}
\node[blue] at (345:0.375) {\large $i$};
\node[blue] at (15:1.25) {\large $j_1$};
\node[blue] at (75:1.25) {\large $j_2$};
\node[blue] at (100:1.25) {\large $j_3$};
\node[blue] at (160:1.25) {\large $j_4$};
\node[blue] at (255:1.25) {\large $j_5$};
\node[blue] at (325:1.25) {\large $j_6$};
\end{scope}

\end{tikzpicture}
\caption{Left: spanning tree contributing to $X^{\G}_4(i)$ on the triangular lattice $\G=\mathcal{L}_{\text{T}}$, with only the paths between $i$, its neighbors and the sink (sent to infinity) drawn. Right: the corresponding unoriented spanning forest of the type $\sigma=i|j_1 j_6|j_3 j_4|j_2 j_5 s$ obtained by removing the edges between $i$ and its neighbors.}
\label{tree_forest}
\end{figure}

Let us now return to the main point of interest of this paper, namely sandpile probabilities. As recalled in Eq.~\eqref{P_X}, one can evaluate them by computing the fractions of spanning trees on the extended graph $\G^*$ such that the reference site $i$ has a fixed number of predecessors among its neighbors. The connection with spanning forests with nodes goes as follows. Let $\mathcal{T}$ be a spanning tree on $\G^*$ such that $i$ has $0\le q\le\deg_{\G^*}(i){-}1$ predecessors among its nearest neighbors, which we denote by $j_1,\ldots,j_q$. Let $\mathcal{F}$ be the spanning forest obtained by removing all the edges of $\mathcal{T}$ that contain $i$, as illustrated in Fig.~\ref{tree_forest}. The vertex $i$, its neighbors and the sink are taken as the nodes, and are distributed in trees of $\mathcal{F}$ according to a partition $\sigma$, with the following properties:
\vspace{-3mm}
\begin{itemize}
\item[$\bullet$] $i$ is an isolated vertex,
\item[$\bullet$] the neighbors of $i$ not in $\{j_1,\ldots,j_q\}$ lie on a single tree connected to the sink $s$,
\item[$\bullet$] $j_1,\ldots,j_q$ lie on (possibly distinct) trees disconnected from the sink.
\end{itemize}
\vspace{-2mm}
Any spanning tree $\mathcal{T}'$ possessing the same predecessorship relations between $i$ and its neighbors as $\mathcal{T}$ will give rise to the same partition: $\sigma(\mathcal{T})=\sigma(\mathcal{T}')$. Conversely any forest on $\G^*$ of the type $\sigma$ that meets these requirements can be extended into a spanning tree contributing to $X^{\G}_q(i)$ by adding a finite number of edges between $i$ and its neighbors (note that the choice of these extra edges is not necessarily unique). We can therefore use the result \eqref{grove_thm} to compute the fractions $X^{\G}_q(i)$, provided we slightly modify the graph by removing a few edges, so that $i$ and its neighbors lie around a single face (to form an annular-one surface graph).


\section{The triangular lattice}
\label{sec4}
\begin{figure}[b]
\centering
\begin{tikzpicture}[scale=.8,blue]
\clip (-4,-3) rectangle (4,3);
\foreach \x in {-5,-4,...,5}{
\draw[help lines] (-5,{\x*sqrt(3)/2})--(5,{\x*sqrt(3)/2});
\draw[xshift=\x cm,help lines] (60:6)--(240:6);
\draw[xshift=\x cm,help lines] (120:6)--(300:6);
}
\draw[dashed,thick] (-5,0)--(5,0);
\draw[dashed,thick] (120:6)--(300:6);
\draw[very thick,->] (0,0)--(0:1);
\draw[very thick,->] (0,0)--(120:1);
\node at (3.5,-0.25) {$x$};
\node at (-2,2.75) {$y$};
\node at (0.5,-0.25) {$\vec{e}_1$};
\node at (-0.5,0.25) {$\vec{e}_2$};
\end{tikzpicture}
\caption{Coordinate system on the triangular lattice.}
\label{tri_coord}
\end{figure}
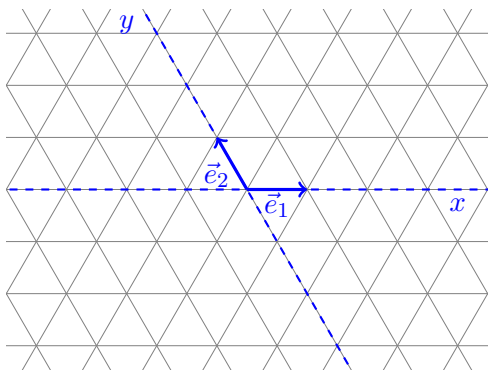

We first consider the Abelian sandpile model on the triangular lattice $\G=\mathcal{L}_{\text{T}}$, and compute one-site probabilities using Eq.~\eqref{P_X}. We associate with each site $i\in\G$ a height $h_i\in\{1,2,3,4,5,6\}$ and coordinates $\vec{r}=(x,y)=x\,\vec{e}_1+y\,\vec{e}_2$, where the unit vectors $\vec{e}_1,\vec{e}_2$ form an angle of $120^\circ$ (see Fig.~\ref{tri_coord}). The six neighbors of the origin $(0,0)$ are therefore located at $(1,0)$, $(1,1)$, $(0,1)$, $(-1,0)$, $(-1,-1)$ and $(0,-1)$. The Euclidean distance $r$ separating $\vec{r}=(x,y)$ from the origin $(0,0)$ is given by $r^2=x^2+y^2-xy$.


\subsection{On the plane}
\begin{figure}[t]
\centering
\begin{tikzpicture}[scale=0.8]
\tikzstyle arrowstyle=[scale=1]
\tikzstyle directed=[postaction={decorate,decoration={markings,mark=at position 0.5 with {\arrow[arrowstyle]{stealth}}}}]

\begin{scope}[xshift=0cm]
\clip (-4,-3) rectangle (4,3);
\foreach \x in {-5,-4,...,5}{
\draw[help lines] (-5,{\x*sqrt(3)/2})--(5,{\x*sqrt(3)/2});
\draw[xshift=\x cm,help lines] (60:6)--(240:6);
\draw[xshift=\x cm,help lines] (120:6)--(300:6);
}
\filldraw (0,0) circle (0.075cm) node[below left] {0};
\filldraw [red](30:{sqrt(3)/3} ) circle (0.075cm);
\draw[very thick,red] ++(30:{sqrt(3)/3} )--++(-60:5);
\filldraw (0,0) circle (0.075cm) node[below left] {0};
\foreach \x in {0,1,...,3}{
\draw[dashed,thick,directed,blue] ({0.5*\x},{-\x*sqrt(3)/2})--({0.5*\x+1},{-\x*sqrt(3)/2});
\draw[dashed,thick,directed,blue] ({0.5*(1+\x)},{-(1+\x)*sqrt(3)/2})--({0.5*\x+1},{-\x*sqrt(3)/2});}
\end{scope}

\begin{scope}[xshift=10cm]
\clip (-4,-3) rectangle (4,3);
\foreach \x in {-5,-4,...,5}{
\draw[help lines] (-5,{\x*sqrt(3)/2})--(5,{\x*sqrt(3)/2});
\draw[xshift=\x cm,help lines] (60:6)--(240:6);
\draw[xshift=\x cm,help lines] (120:6)--(300:6);
}
\filldraw[blue!15!white] (0,0)--(0:1)--(60:1)--(120:1)--(180:1)--(240:1)--(300:1)--(0,0);
\draw[very thick,blue] (0,0)--(0:1)--(60:1)--(120:1)--(180:1)--(240:1)--(300:1)--(0,0);
\filldraw[blue] (0,0) circle (0.075cm) node[left] {${7}$};
\filldraw[blue] (0:1) circle (0.075cm) node[above right] {${1}$};
\filldraw[blue] (60:1) circle (0.075cm) node[above right] {${2}$};
\filldraw[blue] (120:1) circle (0.075cm) node[above left] {${3}$};
\filldraw[blue] (180:1) circle (0.075cm) node[left] {${4}$};
\filldraw[blue] (240:1) circle (0.075cm) node[below left] {${5}$};
\filldraw[blue] (300:1) circle (0.075cm) node[below left] {${6}$};
\filldraw[red] (0.5,{sqrt(3)/6}) circle (0.075cm);
\draw[very thick,red] ++(30:{sqrt(3)/3} )--++(-60:5);
\draw[very thick,->,red] (0.875,{-sqrt(3)/2})--(1.375,{-sqrt(3)/2});
\node[blue] at (3,2.5) {$\mathbf{8=\infty}$};
\node[blue] at (-2.75,-2.25) {$\mathbf{i{=}7{=}(0,0)}$};
\end{scope}

\end{tikzpicture}
\caption{Left: zipper line on the triangular lattice, and zipper edges $(k,\l)$ with nontrivial parallel transport $\phi_{k,\ell}=z$. Right: the modified graph $\xbar{\G}$ obtained by cutting edges between node 7 and its neighbors 2,3,4,5. Node 8 corresponds to the sink/root, and will eventually be sent to infinity in sandpile computations. The zipper extends down to infinity.}
\label{tri_zip}
\end{figure}
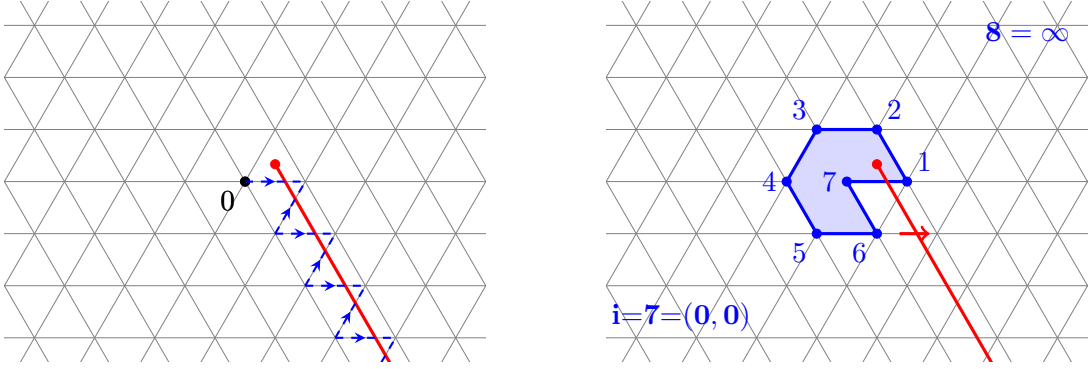

The standard graph Laplacian $\Delta$ on $\G$ with a sink $s$ is defined by Eq.~\eqref{top_mat}, with each vertex having degree six on the lattice. Due to translation invariance, the standard Green function $G_{u,v}=\left(\Delta^{-1}\right)_{u,v}$ only depends on the difference $v-u\equiv(x,y)$. Its Fourier representation reads
\begin{equation}
G_{u,v}=G(v-u)\equiv G(x,y)=\int_{-\pi}^{\pi}\frac{\diff\theta_1}{2\pi}\int_{-\pi}^{\pi}\frac{\diff\theta_2}{2\pi}\frac{e^{\i x\theta_1+\i y\theta_2}}{6-2\cos\theta_1-2\cos\theta_2-2\cos(\theta_1+\theta_2)}.
\label{tri_gf}
\end{equation}
Due to the symmetries of the lattice, the Green function satisfies twelve identities, which can be obtained by repeated applications of the two following relations,
\begin{equation}
G(x,y)=G(x-y,x)=G(x-y,-y),
\end{equation}
corresponding respectively to a counterclockwise rotation of $60^\circ$ and a reflection with respect to the horizontal axis. Although the integral \eqref{tri_gf} diverges, the difference $G(x,y){-}G(0,0)$ is finite. Its values for short distances, as well as its asymptotic behavior, are collected in the appendix. To compute probabilities for predecessor diagrams, we use a line bundle Laplacian $\mathbf{\Delta}$ with a zipper attached to the face whose lower left corner is the origin (see the left panel of Fig.~\ref{tri_zip}). We choose here a nontrivial parallel transport $z$ on the edges of the form $((0,k),(1,k))$ and $((0,k{-}1),(1,k))$ for $k\le 0$.

The reference site $i$ for the fractions $X_q(i)$ is taken to be the origin without loss of generality. The site $i$, its six neighbors and the sink form the subset of selected vertices called nodes (see Section~\ref{sec3}). To meet the requirement that all nodes but the sink lie along the boundary of a single face $f$, we define a modified graph $\xbar{\G}$ by cutting the edges between $i$ and four of its neighbors, as depicted on the right panel of  Fig.~\ref{tri_zip}. We label $i$ and its neighbors from 1 to 7 in counterclockwise order along the boundary of $f$, starting at the right of the zipper.

The fraction $X_0(i)$ is obtained using Eq.~\eqref{X0_comp} and can be written in terms of the standard Green function \eqref{tri_gf} evaluated at $i$ and its neighbors. Its explicit value immediately yields the height-one probability on the triangular lattice:
\begin{equation}
\P_1(i)=\frac{X_{0}(i)}{6}=-\frac{25}{648}-\frac{55}{72\sqrt{3}\pi}+\frac{7}{3\pi^2}+\frac{11\sqrt{3}}{\pi^3}-\frac{90}{\pi^4}+\frac{54\sqrt{3}}{\pi^5}\simeq 0.054.
\end{equation}

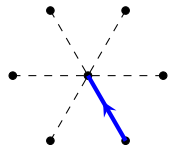
\begin{figure}[b]
\centering
\begin{tikzpicture}
\tikzstyle arrowstyle=[scale=0.9]
\tikzstyle directed=[postaction={decorate,decoration={markings,mark=at position 0.6 with {\arrow[arrowstyle]{stealth}}}}]
\draw[dashed] (0:1)--(180:1);
\draw[dashed] (60:1)--(240:1);
\draw[dashed] (120:1)--(300:1);
\filldraw (0,0) circle (0.05cm);
\filldraw (0:1) circle (0.05cm);
\filldraw (60:1) circle (0.05cm);
\filldraw (120:1) circle (0.05cm);
\filldraw (180:1) circle (0.05cm);
\filldraw (240:1) circle (0.05cm);
\filldraw (300:1) circle (0.05cm);
\draw[ultra thick,directed,blue] (300:1)--(0,0);
\end{tikzpicture}
\caption{Schematic representation of the unique diagram (up to rotations of $60^\circ$) contributing to $X_1(i)$ on the triangular lattice. The isolated dots represent neighbors of $i$ that are not its predecessors.}
\label{tri_X1}
\end{figure}

Next we compute the fraction of spanning trees on $\G^*$ such that $i$ has exactly one predecessor among its nearest neighbors (see Fig.~\ref{tri_X1}), which we assume to be node 6 at $(0,-1)$. By removing the edge between nodes 6 and 7, we can establish a bijection between the spanning trees with a unique predecessor of $i$ to two-component spanning forests, due to the relative positions of the nodes $1,6,7$ on $\xbar{\G}$:
\begin{equation}
X_1(i)=6\times 5\times\frac{\xbar{Z}[6|1234578]}{Z} = 30 \times\frac{\xbar{Z}[6|578]}{Z},
\label{tri_X1_comp}
\end{equation}
where the combinatorial factor takes into account the number of choices for the unique predecessor of $i$ among its neighbors (6) and the number of ways to connect $i$ to the tree with the sink (5). The second equality follows because, $\xbar{\G}$ being planar, the nodes 1 to 4 necessarily belong to the same component as 5,7,8. Considering the spanning forests in $\xbar{Z}[6|578]$, we can write
\begin{equation}
\xbar{Z}[6|578] = \xbar{Z}[6|78] - \xbar{Z}[56|78],
\end{equation}
since node 5 can either be with node 6 or with nodes 7 and 8, implying $\xbar{Z}[6|78] = \xbar{Z}[6|578] + \xbar{Z}[56|78]$.


The number of spanning forests of the types $6|78$ and $56|78$ can be expressed in terms of the usual Green function $\xbar{G}=\xbar{\Gr}|_{z=1}$ and the Green function derivative $\xbar{G}^{\,\prime}=\diff\,\xbar{\Gr}/\diff z|_{z=1}$, where the bar is a reminder that these quantities are defined on the modified graph $\xbar{\G}$. We can compute the number of forests of the first type directly with the all-minors matrix-tree theorem, as the sum on the right-hand side of Eq.~\eqref{AMMT} contains only one term:
\begin{equation}
\xbar{Z}[6|78]=\xbar{Z}\det\xbar{G}_{\{6,7\}}^{\{6,8\}}=\xbar{Z}\big(\xbar{G}_{6,6}\,\xbar{G}_{7,8}-\xbar{G}_{6,8}\,\xbar{G}_{7,6}\big)=\xbar{Z}\big(\xbar{G}_{6,6}-\xbar{G}_{6,7}\big),
\end{equation}
where we used $\xbar{G}_{i,8}=1$ since node 8 is the sink, and $\xbar{G}_{j,i}=\xbar{G}_{i,j}$ for an undirected graph.

The number of spanning forests of the type $56|78$, on the other hand, is harder to compute. It requires the use of the Green function derivative $\xbar{G}^{\,\prime}$ associated with the zipper. Using the generalization of the all-minors matrix-tree theorem to graphs with a connection, one finds the following result (see Section 5.4 in \cite{KW15}):
\begin{equation}
\begin{split}
\xbar{Z}[56|78]&=\lim_{z\to 1}\xbar{\Zr}[56|78]=\lim_{z\to 1}\xbar{\Zr}\times\frac{\det\xbar{\Gr}_{\{5,7\}}^{\{6,8\}}-\det\xbar{\Gr}_{\{5,6\}}^{\{7,8\}}-z^2\det\xbar{\Gr}_{\{6,7\}}^{\{5,8\}}}{1-z^2}\\
&=\xbar{Z}\times\frac{\xbar{G}^{\,\prime}_{5,6}-\xbar{G}^{\,\prime}_{7,6}-\xbar{G}^{\,\prime}_{5,7}+\xbar{G}^{\,\prime}_{6,7}-2\,\xbar{G}_{6,5}+2\,\xbar{G}_{7,5}-\xbar{G}^{\,\prime}_{6,5}+\xbar{G}^{\,\prime}_{7,5}}{-2}\\
&=\xbar{Z}\big(\xbar{G}_{5,6}-\xbar{G}_{5,7}-\xbar{G}^{\,\prime}_{5,6}+\xbar{G}^{\,\prime}_{5,7}-\xbar{G}^{\,\prime}_{6,7}\big),
\end{split}
\label{ztoone}
\end{equation}
where the antisymmetry $\xbar{G}^{\,\prime}_{i,j} = -\xbar{G}^{\,\prime}_{j,i}$ has been used, see \eqref{gp_def}.

Both $\xbar{G},\xbar{G}^{\,\prime}$ on the modified graph $\xbar{\G}$, as well as the ratio $\xbar{Z}/Z$, are given in terms of $G,G'$ on the original graph $\G$ using the Woodbury formula, which we recall in the appendix along with selected values of $G,G'$. The resulting analytical value of $X_1(i)$ on the triangular lattice $\G=\mathcal{L}_{\text{T}}$ reads:
\begin{equation}
X_1(i)=\frac{485}{1296}+\frac{2395}{36\sqrt{3}\pi}-\frac{345}{2\pi^2}-\frac{200\sqrt{3}}{\pi^3}+\frac{2475}{\pi^4}-\frac{1620\sqrt{3}}{\pi^5}\simeq 0.190,
\end{equation}
which, upon using Eq.~\eqref{P_X}, directly yields the height-two probability $\P_2(i)$, shown below.

\begin{figure}[t]
\centering
\begin{tikzpicture}[scale=0.6]
\tikzstyle arrowstyle=[scale=1]
\tikzstyle directed=[postaction={decorate,decoration={markings,mark=at position 0.6 with {\arrow[arrowstyle]{stealth}}}}]

\begin{scope}[xshift=0cm,yshift=-3cm]
\draw[dashed] (0:1)--(180:1);
\draw[dashed] (60:1)--(240:1);
\draw[dashed] (120:1)--(300:1);
\filldraw (0,0) circle (0.05cm);
\filldraw (0:1) circle (0.05cm);
\filldraw (60:1) circle (0.05cm);
\filldraw (120:1) circle (0.05cm);
\filldraw (180:1) circle (0.05cm);
\filldraw (240:1) circle (0.05cm);
\filldraw (300:1) circle (0.05cm);
\draw[ultra thick,directed,blue] (180:1)--(0,0);
\draw[ultra thick,directed,blue] (120:1)--(0,0);
\node at (-2,0) {\large $X_2$};
\end{scope}

\begin{scope}[xshift=3cm,yshift=-3cm]
\draw[dashed] (0:1)--(180:1);
\draw[dashed] (60:1)--(240:1);
\draw[dashed] (120:1)--(300:1);
\filldraw (0,0) circle (0.05cm);
\filldraw (0:1) circle (0.05cm);
\filldraw (60:1) circle (0.05cm);
\filldraw (120:1) circle (0.05cm);
\filldraw (180:1) circle (0.05cm);
\filldraw (240:1) circle (0.05cm);
\filldraw (300:1) circle (0.05cm);
\draw[ultra thick,directed,blue] (180:1)--(0,0);
\draw[ultra thick,directed,blue] (60:1)--(0,0);
\end{scope}

\begin{scope}[xshift=6cm,yshift=-3cm]
\draw[dashed] (0:1)--(180:1);
\draw[dashed] (60:1)--(240:1);
\draw[dashed] (120:1)--(300:1);
\filldraw (0,0) circle (0.05cm);
\filldraw (0:1) circle (0.05cm);
\filldraw (60:1) circle (0.05cm);
\filldraw (120:1) circle (0.05cm);
\filldraw (180:1) circle (0.05cm);
\filldraw (240:1) circle (0.05cm);
\filldraw (300:1) circle (0.05cm);
\draw[ultra thick,directed,blue] (180:1)--(0,0);
\draw[ultra thick,directed,blue] (0:1)--(0,0);
\end{scope}

\begin{scope}[xshift=9cm,yshift=-3cm]
\draw[dashed] (0:1)--(180:1);
\draw[dashed] (60:1)--(240:1);
\draw[dashed] (120:1)--(300:1);
\filldraw (0,0) circle (0.05cm);
\filldraw (0:1) circle (0.05cm);
\filldraw (60:1) circle (0.05cm);
\filldraw (120:1) circle (0.05cm);
\filldraw (180:1) circle (0.05cm);
\filldraw (240:1) circle (0.05cm);
\filldraw (300:1) circle (0.05cm);
\draw[ultra thick,directed,blue] (120:1)to[out=180,in=160](180:1);
\draw[ultra thick,directed,blue] (180:1)--(0,0);
\end{scope}

\begin{scope}[xshift=0cm,yshift=-6cm]
\draw[dashed] (0:1)--(180:1);
\draw[dashed] (60:1)--(240:1);
\draw[dashed] (120:1)--(300:1);
\filldraw (0,0) circle (0.05cm);
\filldraw (0:1) circle (0.05cm);
\filldraw (60:1) circle (0.05cm);
\filldraw (120:1) circle (0.05cm);
\filldraw (180:1) circle (0.05cm);
\filldraw (240:1) circle (0.05cm);
\filldraw (300:1) circle (0.05cm);
\draw[ultra thick,directed,blue] (180:1)--(0,0);
\draw[ultra thick,directed,blue] (120:1)--(0,0);
\draw[ultra thick,directed,blue] (60:1)--(0,0);
\node at (-2,0) {\large $X_3$};
\end{scope}

\begin{scope}[xshift=3cm,yshift=-6cm]
\draw[dashed] (0:1)--(180:1);
\draw[dashed] (60:1)--(240:1);
\draw[dashed] (120:1)--(300:1);
\filldraw (0,0) circle (0.05cm);
\filldraw (0:1) circle (0.05cm);
\filldraw (60:1) circle (0.05cm);
\filldraw (120:1) circle (0.05cm);
\filldraw (180:1) circle (0.05cm);
\filldraw (240:1) circle (0.05cm);
\filldraw (300:1) circle (0.05cm);
\draw[ultra thick,directed,blue] (180:1)--(0,0);
\draw[ultra thick,directed,blue] (120:1)--(0,0);
\draw[ultra thick,directed,blue] (0:1)--(0,0);
\end{scope}

\begin{scope}[xshift=6cm,yshift=-6cm]
\draw[dashed] (0:1)--(180:1);
\draw[dashed] (60:1)--(240:1);
\draw[dashed] (120:1)--(300:1);
\filldraw (0,0) circle (0.05cm);
\filldraw (0:1) circle (0.05cm);
\filldraw (60:1) circle (0.05cm);
\filldraw (120:1) circle (0.05cm);
\filldraw (180:1) circle (0.05cm);
\filldraw (240:1) circle (0.05cm);
\filldraw (300:1) circle (0.05cm);
\draw[ultra thick,directed,blue] (180:1)--(0,0);
\draw[ultra thick,directed,blue] (60:1)--(0,0);
\draw[ultra thick,directed,blue] (300:1)--(0,0);
\end{scope}

\begin{scope}[xshift=9cm,yshift=-6cm]
\draw[dashed] (0:1)--(180:1);
\draw[dashed] (60:1)--(240:1);
\draw[dashed] (120:1)--(300:1);
\filldraw (0,0) circle (0.05cm);
\filldraw (0:1) circle (0.05cm);
\filldraw (60:1) circle (0.05cm);
\filldraw (120:1) circle (0.05cm);
\filldraw (180:1) circle (0.05cm);
\filldraw (240:1) circle (0.05cm);
\filldraw (300:1) circle (0.05cm);
\draw[ultra thick,directed,blue] (120:1)to[out=180,in=160](180:1);
\draw[ultra thick,directed,blue] (180:1)--(0,0);
\draw[ultra thick,directed,blue] (60:1)--(0,0);
\end{scope}

\begin{scope}[xshift=12cm,yshift=-6cm]
\draw[dashed] (0:1)--(180:1);
\draw[dashed] (60:1)--(240:1);
\draw[dashed] (120:1)--(300:1);
\filldraw (0,0) circle (0.05cm);
\filldraw (0:1) circle (0.05cm);
\filldraw (60:1) circle (0.05cm);
\filldraw (120:1) circle (0.05cm);
\filldraw (180:1) circle (0.05cm);
\filldraw (240:1) circle (0.05cm);
\filldraw (300:1) circle (0.05cm);
\draw[ultra thick,directed,blue] (120:1)to[out=180,in=160](180:1);
\draw[ultra thick,directed,blue] (180:1)--(0,0);
\draw[ultra thick,directed,blue] (0:1)--(0,0);
\end{scope}

\begin{scope}[xshift=15cm,yshift=-6cm]
\draw[dashed] (0:1)--(180:1);
\draw[dashed] (60:1)--(240:1);
\draw[dashed] (120:1)--(300:1);
\filldraw (0,0) circle (0.05cm);
\filldraw (0:1) circle (0.05cm);
\filldraw (60:1) circle (0.05cm);
\filldraw (120:1) circle (0.05cm);
\filldraw (180:1) circle (0.05cm);
\filldraw (240:1) circle (0.05cm);
\filldraw (300:1) circle (0.05cm);
\draw[ultra thick,directed,blue] (60:1)to[out=100,in=20](120:1.5)to[out=200,in=170](180:1);
\draw[ultra thick,directed,blue] (180:1)--(0,0);
\draw[ultra thick,directed,blue] (120:1)--(0,0);
\end{scope}

\begin{scope}[xshift=18cm,yshift=-6cm]
\draw[dashed] (0:1)--(180:1);
\draw[dashed] (60:1)--(240:1);
\draw[dashed] (120:1)--(300:1);
\filldraw (0,0) circle (0.05cm);
\filldraw (0:1) circle (0.05cm);
\filldraw (60:1) circle (0.05cm);
\filldraw (120:1) circle (0.05cm);
\filldraw (180:1) circle (0.05cm);
\filldraw (240:1) circle (0.05cm);
\filldraw (300:1) circle (0.05cm);
\draw[ultra thick,directed,blue] (60:1)to[out=100,in=20](120:1.5)to[out=200,in=170](180:1);
\draw[ultra thick,directed,blue] (180:1)--(0,0);
\draw[ultra thick,directed,blue] (120:1)to(120:1.5);
\end{scope}

\begin{scope}[xshift=0cm,yshift=-9cm]
\draw[dashed] (0:1)--(180:1);
\draw[dashed] (60:1)--(240:1);
\draw[dashed] (120:1)--(300:1);
\filldraw (0,0) circle (0.05cm);
\filldraw (0:1) circle (0.05cm);
\filldraw (60:1) circle (0.05cm);
\filldraw (120:1) circle (0.05cm);
\filldraw (180:1) circle (0.05cm);
\filldraw (240:1) circle (0.05cm);
\filldraw (300:1) circle (0.05cm);
\draw[ultra thick,directed,blue] (180:1)--(0,0);
\draw[ultra thick,directed,blue] (120:1)--(0,0);
\draw[ultra thick,directed,blue] (60:1)--(0,0);
\draw[ultra thick,directed,blue] (0:1)--(0,0);
\node at (-2,0) {\large $X_4$};
\end{scope}

\begin{scope}[xshift=3cm,yshift=-9cm]
\draw[dashed] (0:1)--(180:1);
\draw[dashed] (60:1)--(240:1);
\draw[dashed] (120:1)--(300:1);
\filldraw (0,0) circle (0.05cm);
\filldraw (0:1) circle (0.05cm);
\filldraw (60:1) circle (0.05cm);
\filldraw (120:1) circle (0.05cm);
\filldraw (180:1) circle (0.05cm);
\filldraw (240:1) circle (0.05cm);
\filldraw (300:1) circle (0.05cm);
\draw[ultra thick,directed,blue] (180:1)--(0,0);
\draw[ultra thick,directed,blue] (120:1)--(0,0);
\draw[ultra thick,directed,blue] (60:1)--(0,0);
\draw[ultra thick,directed,blue] (300:1)--(0,0);
\end{scope}

\begin{scope}[xshift=6cm,yshift=-9cm]
\draw[dashed] (0:1)--(180:1);
\draw[dashed] (60:1)--(240:1);
\draw[dashed] (120:1)--(300:1);
\filldraw (0,0) circle (0.05cm);
\filldraw (0:1) circle (0.05cm);
\filldraw (60:1) circle (0.05cm);
\filldraw (120:1) circle (0.05cm);
\filldraw (180:1) circle (0.05cm);
\filldraw (240:1) circle (0.05cm);
\filldraw (300:1) circle (0.05cm);
\draw[ultra thick,directed,blue] (180:1)--(0,0);
\draw[ultra thick,directed,blue] (120:1)--(0,0);
\draw[ultra thick,directed,blue] (0:1)--(0,0);
\draw[ultra thick,directed,blue] (300:1)--(0,0);
\end{scope}

\begin{scope}[xshift=9cm,yshift=-9cm]
\draw[dashed] (0:1)--(180:1);
\draw[dashed] (60:1)--(240:1);
\draw[dashed] (120:1)--(300:1);
\filldraw (0,0) circle (0.05cm);
\filldraw (0:1) circle (0.05cm);
\filldraw (60:1) circle (0.05cm);
\filldraw (120:1) circle (0.05cm);
\filldraw (180:1) circle (0.05cm);
\filldraw (240:1) circle (0.05cm);
\filldraw (300:1) circle (0.05cm);
\draw[ultra thick,directed,blue] (120:1)to[out=180,in=160](180:1);
\draw[ultra thick,directed,blue] (180:1)--(0,0);
\draw[ultra thick,directed,blue] (60:1)--(0,0);
\draw[ultra thick,directed,blue] (0:1)--(0,0);
\end{scope}

\begin{scope}[xshift=12cm,yshift=-9cm]
\draw[dashed] (0:1)--(180:1);
\draw[dashed] (60:1)--(240:1);
\draw[dashed] (120:1)--(300:1);
\filldraw (0,0) circle (0.05cm);
\filldraw (0:1) circle (0.05cm);
\filldraw (60:1) circle (0.05cm);
\filldraw (120:1) circle (0.05cm);
\filldraw (180:1) circle (0.05cm);
\filldraw (240:1) circle (0.05cm);
\filldraw (300:1) circle (0.05cm);
\draw[ultra thick,directed,blue] (120:1)to[out=180,in=160](180:1);
\draw[ultra thick,directed,blue] (180:1)--(0,0);
\draw[ultra thick,directed,blue] (0:1)--(0,0);
\draw[ultra thick,directed,blue] (300:1)--(0,0);
\end{scope}

\begin{scope}[xshift=15cm,yshift=-9cm]
\draw[dashed] (0:1)--(180:1);
\draw[dashed] (60:1)--(240:1);
\draw[dashed] (120:1)--(300:1);
\filldraw (0,0) circle (0.05cm);
\filldraw (0:1) circle (0.05cm);
\filldraw (60:1) circle (0.05cm);
\filldraw (120:1) circle (0.05cm);
\filldraw (180:1) circle (0.05cm);
\filldraw (240:1) circle (0.05cm);
\filldraw (300:1) circle (0.05cm);
\draw[ultra thick,directed,blue] (120:1)to[out=180,in=160](180:1);
\draw[ultra thick,directed,blue] (180:1)--(0,0);
\draw[ultra thick,directed,blue] (60:1)--(0,0);
\draw[ultra thick,directed,blue] (300:1)--(0,0);
\end{scope}

\begin{scope}[xshift=0cm,yshift=-12cm]
\draw[dashed] (0:1)--(180:1);
\draw[dashed] (60:1)--(240:1);
\draw[dashed] (120:1)--(300:1);
\filldraw (0,0) circle (0.05cm);
\filldraw (0:1) circle (0.05cm);
\filldraw (60:1) circle (0.05cm);
\filldraw (120:1) circle (0.05cm);
\filldraw (180:1) circle (0.05cm);
\filldraw (240:1) circle (0.05cm);
\filldraw (300:1) circle (0.05cm);
\draw[ultra thick,directed,blue] (120:1)to[out=180,in=160](180:1);
\draw[ultra thick,directed,blue] (180:1)--(0,0);
\draw[ultra thick,directed,blue] (60:1)--(0,0);
\draw[ultra thick,directed,blue] (240:1)--(0,0);
\end{scope}

\begin{scope}[xshift=3cm,yshift=-12cm]
\draw[dashed] (0:1)--(180:1);
\draw[dashed] (60:1)--(240:1);
\draw[dashed] (120:1)--(300:1);
\filldraw (0,0) circle (0.05cm);
\filldraw (0:1) circle (0.05cm);
\filldraw (60:1) circle (0.05cm);
\filldraw (120:1) circle (0.05cm);
\filldraw (180:1) circle (0.05cm);
\filldraw (240:1) circle (0.05cm);
\filldraw (300:1) circle (0.05cm);
\draw[ultra thick,directed,blue] (60:1)to[out=100,in=20](120:1.5)to[out=200,in=170](180:1);
\draw[ultra thick,directed,blue] (180:1)--(0,0);
\draw[ultra thick,directed,blue] (120:1)--(0,0);
\draw[ultra thick,directed,blue] (0:1)--(0,0);
\end{scope}

\begin{scope}[xshift=6cm,yshift=-12cm]
\draw[dashed] (0:1)--(180:1);
\draw[dashed] (60:1)--(240:1);
\draw[dashed] (120:1)--(300:1);
\filldraw (0,0) circle (0.05cm);
\filldraw (0:1) circle (0.05cm);
\filldraw (60:1) circle (0.05cm);
\filldraw (120:1) circle (0.05cm);
\filldraw (180:1) circle (0.05cm);
\filldraw (240:1) circle (0.05cm);
\filldraw (300:1) circle (0.05cm);
\draw[ultra thick,directed,blue] (60:1)to[out=100,in=20](120:1.5)to[out=200,in=170](180:1);
\draw[ultra thick,directed,blue] (180:1)--(0,0);
\draw[ultra thick,directed,blue] (120:1)--(0,0);
\draw[ultra thick,directed,blue] (300:1)--(0,0);
\end{scope}

\begin{scope}[xshift=9cm,yshift=-12cm]
\draw[dashed] (0:1)--(180:1);
\draw[dashed] (60:1)--(240:1);
\draw[dashed] (120:1)--(300:1);
\filldraw (0,0) circle (0.05cm);
\filldraw (0:1) circle (0.05cm);
\filldraw (60:1) circle (0.05cm);
\filldraw (120:1) circle (0.05cm);
\filldraw (180:1) circle (0.05cm);
\filldraw (240:1) circle (0.05cm);
\filldraw (300:1) circle (0.05cm);
\draw[ultra thick,directed,blue] (60:1)to[out=100,in=20](120:1.5)to[out=200,in=170](180:1);
\draw[ultra thick,directed,blue] (120:1)to(120:1.5);
\draw[ultra thick,directed,blue] (180:1)--(0,0);
\draw[ultra thick,directed,blue] (0:1)--(0,0);
\end{scope}

\begin{scope}[xshift=12cm,yshift=-12cm]
\draw[dashed] (0:1)--(180:1);
\draw[dashed] (60:1)--(240:1);
\draw[dashed] (120:1)--(300:1);
\filldraw (0,0) circle (0.05cm);
\filldraw (0:1) circle (0.05cm);
\filldraw (60:1) circle (0.05cm);
\filldraw (120:1) circle (0.05cm);
\filldraw (180:1) circle (0.05cm);
\filldraw (240:1) circle (0.05cm);
\filldraw (300:1) circle (0.05cm);
\draw[ultra thick,directed,blue] (60:1)to[out=100,in=20](120:1.5)to[out=200,in=170](180:1);
\draw[ultra thick,directed,blue] (120:1)to(120:1.5);
\draw[ultra thick,directed,blue] (180:1)--(0,0);
\draw[ultra thick,directed,blue] (300:1)--(0,0);
\end{scope}

\begin{scope}[xshift=15cm,yshift=-12cm]
\draw[dashed] (0:1)--(180:1);
\draw[dashed] (60:1)--(240:1);
\draw[dashed] (120:1)--(300:1);
\filldraw (0,0) circle (0.05cm);
\filldraw (0:1) circle (0.05cm);
\filldraw (60:1) circle (0.05cm);
\filldraw (120:1) circle (0.05cm);
\filldraw (180:1) circle (0.05cm);
\filldraw (240:1) circle (0.05cm);
\filldraw (300:1) circle (0.05cm);
\draw[ultra thick,directed,blue] (0:1)to[out=20,in=0](90:1.5)to[out=180,in=160](180:1);
\draw[ultra thick,directed,blue] (180:1)--(0,0);
\draw[ultra thick,directed,blue] (120:1)--(0,0);
\draw[ultra thick,directed,blue] (60:1)--(0,0);
\end{scope}

\begin{scope}[xshift=0cm,yshift=-15cm]
\draw[dashed] (0:1)--(180:1);
\draw[dashed] (60:1)--(240:1);
\draw[dashed] (120:1)--(300:1);
\filldraw (0,0) circle (0.05cm);
\filldraw (0:1) circle (0.05cm);
\filldraw (60:1) circle (0.05cm);
\filldraw (120:1) circle (0.05cm);
\filldraw (180:1) circle (0.05cm);
\filldraw (240:1) circle (0.05cm);
\filldraw (300:1) circle (0.05cm);
\draw[ultra thick,directed,blue] (0:1)to[out=20,in=0](90:1.5)to[out=180,in=160](180:1);
\draw[ultra thick,directed,blue] (180:1)--(0,0);
\draw[ultra thick,directed,blue] (120:1)--(0,0);
\draw[ultra thick,directed,blue] (60:1)to[out=160,in=260](90:1.5);
\end{scope}

\begin{scope}[xshift=3cm,yshift=-15cm]
\draw[dashed] (0:1)--(180:1);
\draw[dashed] (60:1)--(240:1);
\draw[dashed] (120:1)--(300:1);
\filldraw (0,0) circle (0.05cm);
\filldraw (0:1) circle (0.05cm);
\filldraw (60:1) circle (0.05cm);
\filldraw (120:1) circle (0.05cm);
\filldraw (180:1) circle (0.05cm);
\filldraw (240:1) circle (0.05cm);
\filldraw (300:1) circle (0.05cm);
\draw[ultra thick,directed,blue] (0:1)to[out=20,in=0](90:1.5)to[out=180,in=160](180:1);
\draw[ultra thick,directed,blue] (180:1)--(0,0);
\draw[ultra thick,directed,blue] (60:1)to[out=150,in=30](120:1);
\draw[ultra thick,directed,blue] (120:1)--(0,0);
\end{scope}

\begin{scope}[xshift=6cm,yshift=-15cm]
\draw[dashed] (0:1)--(180:1);
\draw[dashed] (60:1)--(240:1);
\draw[dashed] (120:1)--(300:1);
\filldraw (0,0) circle (0.05cm);
\filldraw (0:1) circle (0.05cm);
\filldraw (60:1) circle (0.05cm);
\filldraw (120:1) circle (0.05cm);
\filldraw (180:1) circle (0.05cm);
\filldraw (240:1) circle (0.05cm);
\filldraw (300:1) circle (0.05cm);
\draw[ultra thick,directed,blue] (0:1)to[out=20,in=340](60:1.5)to[out=160,in=20](120:1.5)to[out=200,in=160](180:1);
\draw[ultra thick,directed,blue] (180:1)--(0,0);
\draw[ultra thick,directed,blue] (120:1)--(120:1.5);
\draw[ultra thick,directed,blue] (60:1)--(60:1.5);
\end{scope}

\begin{scope}[xshift=9cm,yshift=-15cm]
\draw[dashed] (0:1)--(180:1);
\draw[dashed] (60:1)--(240:1);
\draw[dashed] (120:1)--(300:1);
\filldraw (0,0) circle (0.05cm);
\filldraw (0:1) circle (0.05cm);
\filldraw (60:1) circle (0.05cm);
\filldraw (120:1) circle (0.05cm);
\filldraw (180:1) circle (0.05cm);
\filldraw (240:1) circle (0.05cm);
\filldraw (300:1) circle (0.05cm);
\draw[ultra thick,directed,blue] (120:1)to[out=180,in=160](180:1);
\draw[ultra thick,directed,blue] (180:1)--(0,0);
\draw[ultra thick,directed,blue] (0:1)to[out=20,in=0](60:1);
\draw[ultra thick,directed,blue] (60:1)--(0,0);
\end{scope}

\begin{scope}[xshift=12cm,yshift=-15cm]
\draw[dashed] (0:1)--(180:1);
\draw[dashed] (60:1)--(240:1);
\draw[dashed] (120:1)--(300:1);
\filldraw (0,0) circle (0.05cm);
\filldraw (0:1) circle (0.05cm);
\filldraw (60:1) circle (0.05cm);
\filldraw (120:1) circle (0.05cm);
\filldraw (180:1) circle (0.05cm);
\filldraw (240:1) circle (0.05cm);
\filldraw (300:1) circle (0.05cm);
\draw[ultra thick,directed,blue] (120:1)to[out=180,in=160](180:1);
\draw[ultra thick,directed,blue] (180:1)--(0,0);
\draw[ultra thick,directed,blue] (300:1)to[out=0,in=340](0:1);
\draw[ultra thick,directed,blue] (0:1)--(0,0);
\end{scope}

\end{tikzpicture}
\caption{Classes of predecessor diagrams contributing to $X_q(i)$ for $2\le q\le 4$ on the triangular lattice. A multiplicity (analogous to the factor 30 in Eq.~\eqref{tri_X1_comp}) is associated with each class, accounting for the symmetries of the lattice. The ones for the diagrams of $X_2(i)$, for instance, are given by 24, 24, 12 and 48, respectively.}
\label{tri_diag}
\end{figure}
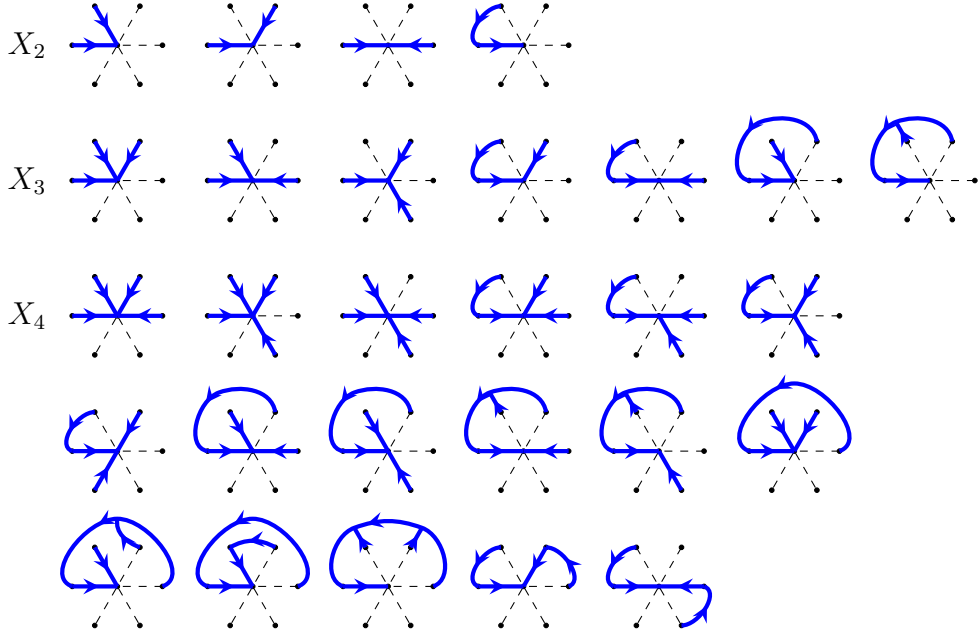

Higher-height probabilities on the lattice are computed in the same way, although there are multiple inequivalent predecessor diagrams contributing to the fractions $X_q(i)$ when $q\ge 2$ (see Fig.~\ref{tri_diag}). The probability of each of these diagrams can be determined similarly to that of the unique one contributing to $X_1(i)$. We find the following explicit expressions for $\P_a(i)$ on $\G=\mathcal{L}_{\text{T}}$:
\begin{subequations}
\begin{align}
\P_1(i)&=-\frac{25}{648}-\frac{55}{72\sqrt{3}\pi}+\frac{7}{3\pi^2}+\frac{11\sqrt{3}}{\pi^3}-\frac{90}{\pi^4}+\frac{54\sqrt{3}}{\pi^5}\simeq 0.054,\\
\P_2(i)&=\frac{47}{1296}+\frac{301}{24\sqrt{3}\pi}-\frac{193}{6\pi^2}-\frac{29\sqrt{3}}{\pi^3}+\frac{405}{\pi^4}-\frac{270\sqrt{3}}{\pi^5}\simeq 0.092,\\
\P_3(i)&=\frac{3}{8}-\frac{5929}{144\sqrt{3}\pi}+\frac{1441}{12\pi^2}-\frac{9\sqrt{3}}{\pi^3}-\frac{720}{\pi^4}+\frac{540\sqrt{3}}{\pi^5}\simeq 0.137,\\
\P_4(i)&=\frac{3427}{2592}+\frac{6515}{144\sqrt{3}\pi}-\frac{2125}{12\pi^2}+\frac{91\sqrt{3}}{\pi^3}+\frac{630}{\pi^4}-\frac{540\sqrt{3}}{\pi^5}\simeq 0.189,\\
\P_5(i)&=-\frac{2663}{1296}-\frac{71\sqrt{3}}{16\pi}+\frac{1331}{12\pi^2}-\frac{94\sqrt{3}}{\pi^3}-\frac{270}{\pi^4}+\frac{270\sqrt{3}}{\pi^5}\simeq 0.242,\\
\P_6(i)&=\frac{1175}{864}-\frac{365}{144\sqrt{3}\pi}-\frac{289}{12\pi^2}+\frac{30\sqrt{3}}{\pi^3}+\frac{45}{\pi^4}-\frac{54\sqrt{3}}{\pi^5}\simeq 0.286,
\end{align}
\label{tri_prob}%
\end{subequations}
where we used $\P_6(i)=1-\sum_{a=1}^{5}\P_a(i)$ to avoid a direct computation of $X_5(i)$, which requires the evaluation of 22 separate predecessor diagrams.

\subsection{On the upper half-plane}

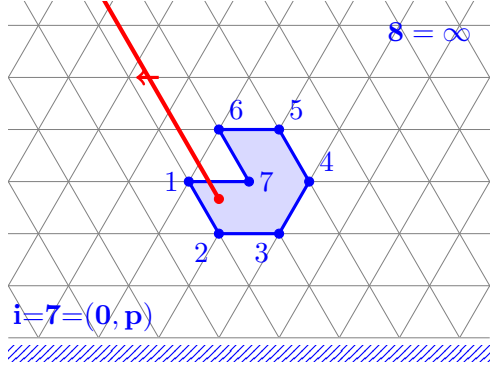
\begin{figure}[t]
\centering
\begin{tikzpicture}[scale=0.8]
\clip (-4,-3) rectangle (4,3);
\foreach \x in {-5,-4,...,5}{
\draw[help lines] (-5,{\x*sqrt(3)/2})--(5,{\x*sqrt(3)/2});
\draw[xshift=\x cm,help lines] (60:6)--(240:6);
\draw[xshift=\x cm,help lines] (120:6)--(300:6);
}
\filldraw[blue!15!white] (0,0)--(120:1)--(60:1)--(0:1)--(300:1)--(240:1)--(180:1)--(0,0);
\draw[very thick,blue] (0,0)--(120:1)--(60:1)--(0:1)--(300:1)--(240:1)--(180:1)--(0,0);
\filldraw[blue] (0,0) circle (0.075cm) node[right] {${7}$};
\filldraw[blue] (0:1) circle (0.075cm) node[above right] {${4}$};
\filldraw[blue] (60:1) circle (0.075cm) node[above right] {${5}$};
\filldraw[blue] (120:1) circle (0.075cm) node[above right] {${6}$};
\filldraw[blue] (180:1) circle (0.075cm) node[left] {${1}$};
\filldraw[blue] (240:1) circle (0.075cm) node[below left] {${2}$};
\filldraw[blue] (300:1) circle (0.075cm) node[below left] {${3}$};
\node[blue] at (-2.75,-2.25) {$\mathbf{i{=}7{=}(0,p)}$};
\filldraw[red] (-0.5,-{sqrt(3)/6}) circle (0.075cm);
\draw[ultra thick,red] ++(210:{sqrt(3)/3} )--++(120:5);
\draw[very thick,->,red] (-1.5,{sqrt(3)})--(-1.875,{sqrt(3)});
\filldraw[white] (-4,{-3*sqrt(3)/2}) rectangle (4,-3);
\draw[help lines] (-4,{-3*sqrt(3)/2})--(4,{-3*sqrt(3)/2});
\fill[pattern=north east lines, pattern color=blue] (-4,{-3*sqrt(3)/2-0.125}) rectangle (4,-3);
\node[blue] at (3,2.5) {$\mathbf{8=\infty}$};
\end{tikzpicture}
\caption{The modified graph $\xbar{\G}$ obtained by cutting edges between node 7 and its neighbors 2,3,4,5 on the triangular half-lattice. The sink corresponds to node 8 and is sent to infinity. The zipper extends up to infinity.}
\label{tri_uhp_cuts}
\end{figure}

In addition to the infinite triangular lattice, we compute one-site probabilities on the semi-infinite lattice with a horizontal boundary, i.e. $\G=\{(x,y)\in\mathcal{L}_{\text{T}}|y>0\}$. Usual boundary conditions on the upper half-plane are either uniformly open or uniformly closed ($\Delta_{i,i}=6$ or $\Delta_{i,i}=4$ resp. for boundary sites). For the latter, we have not been able to define a suitable reflection in order to use the image method. Therefore we only discuss the case of an open boundary, for which a simple reflection through the line $y=0$ works and yields the following Green function,
\begin{equation}
G^{\textrm{op}}_{(x_1,y_1),(x_2,y_2)}=G_{(x_1,y_1),(x_2,y_2)}-G_{(x_1,y_1),(x_2-y_2,-y_2)},
\label{tri_gf_op}
\end{equation}
where $G$ is the standard Green function on the full triangular lattice \eqref{tri_gf}. We choose the reference site $i$ to be located at $(0,p)$ with $p\gg 1$, and take the zipper to be the path on the dual graph crossing the edges of the form $\left((0,k),(-1,k)\right)$ and $\left((0,k{+}1),(-1,k)\right)$ for $k\ge p$ (see Fig.~\ref{tri_uhp_cuts}). For such a zipper, the Green function derivative $G^{\prime\,\mathrm{op}}$ on the upper half-plane reads:
\begin{equation}
\begin{split}
G^{\prime\,\mathrm{op}}_{(x_1,y_1),(x_2,y_2)}&=\sum_{k=0}^{\infty}\Big[G^{\mathrm{op}}_{(x_1,y_1),(-1,p+k)}G^{\mathrm{op}}_{(0,p+k),(x_2,y_2)}-G^{\mathrm{op}}_{(x_1,y_1),(0,p+k)}G^{\mathrm{op}}_{(-1,p+k),(x_2,y_2)}\\
&\quad\quad+G^{\mathrm{op}}_{(x_1,y_1),(-1,p+k)}G^{\mathrm{op}}_{(0,p+k+1),(x_2,y_2)}-G^{\mathrm{op}}_{(x_1,y_1),(0,p+k+1)}G^{\mathrm{op}}_{(-1,p+k),(x_2,y_2)}\Big].
\end{split}
\end{equation}
Using Eq.~\eqref{tri_gf_op}, we can write $G^{\prime\,\mathrm{op}}$ in terms of $G'$ on the full lattice (with respect to the zipper depicted in Fig.~\ref{tri_zip}) as follows:
\begin{equation}
\begin{split}
G^{\prime\,\mathrm{op}}_{(x_1,y_1),(x_2,y_2)}&=G'_{(-x_1,p-y_1),(-x_2,p-y_2)}-G'_{(-x_1,p-y_1),(y_2-x_2,p+y_2)}\\
&\quad-G'_{(y_1-x_1,p+y_1),(-x_2,p-y_2)}+G'_{(y_1-x_1,p+y_1),(y_2-x_2,p+y_2)}.
\end{split}
\label{tri_gp_op}
\end{equation}
For pairs of vertices $(x_i,y_i)$ close to the head of the zipper (i.e. of the form $(a_i,p{+}b_i)$ with $a_i,b_i=o(1)$), the first term of Eq.~\eqref{tri_gp_op} can be computed exactly since it is independent of $p$; the three remaining terms are evaluated as power series in $1/p$ (see the appendix for more details).

Similarly to full-plane computations, the reference site $i$ located at $(0,p)$ and its six neighbors are chosen as nodes (note that we have relabeled the nodes with respect to Fig.~\ref{tri_zip}, so that the new zipper is once again located between nodes 1 and 7). We define a modified graph $\xbar{\G}$ by cutting the edges $\{7,2\}$, $\{7,3\}$, $\{7,4\}$ and $\{7,5\}$, so that nodes 1 to 7 lie along the boundary of a single face on $\xbar{\G}$. Since the upper half-plane is not invariant under rotations of $60^\circ$, there are in total roughly four times as many distinct diagrams as on the full plane (the left-right symmetry is still preserved). The correspondence between predecessor diagrams and spanning forests with a fixed node partition $\sigma$ holds on the upper half-plane as well, with $\xbar{Z}[\sigma]/Z$ given as a function of $G^{\mathrm{op}},G^{\prime\,\mathrm{op}}$ instead of $G,G'$. The final results for one-site probabilities at $i=(0,p)$ on the upper half-plane take the form
\begin{equation}
\sigma^{\text{op}}_a(r)\equiv\P_a^{\mathrm{op}}(r)-\P_a=\frac{1}{r^2}\left(c_a+d_a\log r\right)+\ldots,
\label{tri_uhp_sig}
\end{equation}
where $\P_a$ are the one-site probabilities on the full plane \eqref{tri_prob} and $r=\sqrt{3}\,p/2$ is the Euclidean distance between $i=(0,p)$ and the symmetry axis for the image method, $y=0$. The coefficients $c_a,d_a$ are given in Table~\ref{tri_uhp_coef}, from which we see that, like on the square half-lattice, all one-point height probabilities have a logarithmic term except for the height 1.

\begin{table}[t]
\centering
\large
\renewcommand{\arraystretch}{1.8}
\tabcolsep12pt
\begin{tabular}{|c|p{7.8cm}c|}
\hline
 & \hspace{3cm}$c_a$ & $d_a$\\
\hline
$a=1$ & $-\frac{25}{144\sqrt{3}\pi}-\frac{5}{48\pi^2}+\frac{33\sqrt{3}}{8\pi^3}-\frac{99}{4\pi^4}+\frac{27\sqrt{3}}{2\pi^5}$ & $0$\\
$a=2$ & $\left(-\frac{25}{48\pi^2}+\frac{45\sqrt{3}}{16\pi^3}-\frac{27}{2\pi^4}+\frac{27\sqrt{3}}{4\pi^5}\right)\left(\gamma+\frac{1}{2}\text{{\small $\log 48$}}\right)\newline +\frac{149}{96\sqrt{3}\pi}-\frac{41}{192\pi^2}-\frac{1323\sqrt{3}}{64\pi^3}+\frac{513}{4\pi^4}-\frac{1161\sqrt{3}}{16\pi^5}$ & $-\frac{25}{48\pi^2}+\frac{45\sqrt{3}}{16\pi^3}-\frac{27}{2\pi^4}+\frac{27\sqrt{3}}{4\pi^5}$\\
$a=3$ & $\left(\frac{19}{6\pi^2}-\frac{209\sqrt{3}}{16\pi^3}+\frac{225}{4\pi^4}-\frac{27\sqrt{3}}{\pi^5}\right)\left(\gamma+\frac{1}{2}\text{{\small $\log 48$}}\right)\newline -\frac{353}{72\sqrt{3}\pi}+\frac{275}{48\pi^2}+\frac{2491\sqrt{3}}{64\pi^3}-\frac{4275}{16\pi^4}+\frac{621\sqrt{3}}{4\pi^5}$ & $\frac{19}{6\pi^2}-\frac{209\sqrt{3}}{16\pi^3}+\frac{225}{4\pi^4}-\frac{27\sqrt{3}}{\pi^5}$\\
$a=4$ & $\left(-\frac{469}{96\pi^2}+\frac{341\sqrt{3}}{16\pi^3}-\frac{351}{4\pi^4}+\frac{81\sqrt{3}}{2\pi^5}\right)\left(\gamma+\frac{1}{2}\text{{\small $\log 48$}}\right)\newline +\frac{1019}{144\sqrt{3}\pi}-\frac{5425}{384\pi^2}-\frac{2127\sqrt{3}}{64\pi^3}+\frac{4473}{16\pi^4}-\frac{1323\sqrt{3}}{8\pi^5}$ & $-\frac{469}{96\pi^2}+\frac{341\sqrt{3}}{16\pi^3}-\frac{351}{4\pi^4}+\frac{81\sqrt{3}}{2\pi^5}$\\
$a=5$ & $\left(\frac{167}{48\pi^2}-\frac{235\sqrt{3}}{16\pi^3}+\frac{243}{4\pi^4}-\frac{27\sqrt{3}}{\pi^5}\right)\left(\gamma+\frac{1}{2}\text{{\small $\log 48$}}\right)\newline -\frac{671}{144\sqrt{3}\pi}+\frac{2227}{192\pi^2}+\frac{753\sqrt{3}}{64\pi^3}-\frac{2349}{16\pi^4}+\frac{351\sqrt{3}}{4\pi^5}$ & $\frac{167}{48\pi^2}-\frac{235\sqrt{3}}{16\pi^3}+\frac{243}{4\pi^4}-\frac{27\sqrt{3}}{\pi^5}$\\
$a=6$ & $\left(-\frac{119}{96\pi^2}+\frac{29\sqrt{3}}{8\pi^3}-\frac{63}{4\pi^4}+\frac{27\sqrt{3}}{4\pi^5}\right)\left(\gamma+\frac{1}{2}\text{{\small $\log 48$}}\right)\newline +\frac{319}{288\sqrt{3}\pi}-\frac{369}{128\pi^2}-\frac{29\sqrt{3}}{32\pi^3}+\frac{495}{16\pi^4}-\frac{297\sqrt{3}}{16\pi^5}$ & $-\frac{119}{96\pi^2}+\frac{29\sqrt{3}}{8\pi^3}-\frac{63}{4\pi^4}+\frac{27\sqrt{3}}{4\pi^5}$\\
\hline
\end{tabular}
\caption{Numerical coefficients of one-site probabilities on the triangular upper half-plane.}
\label{tri_uhp_coef}
\end{table}


\section{The hexagonal lattice}
\label{sec5}
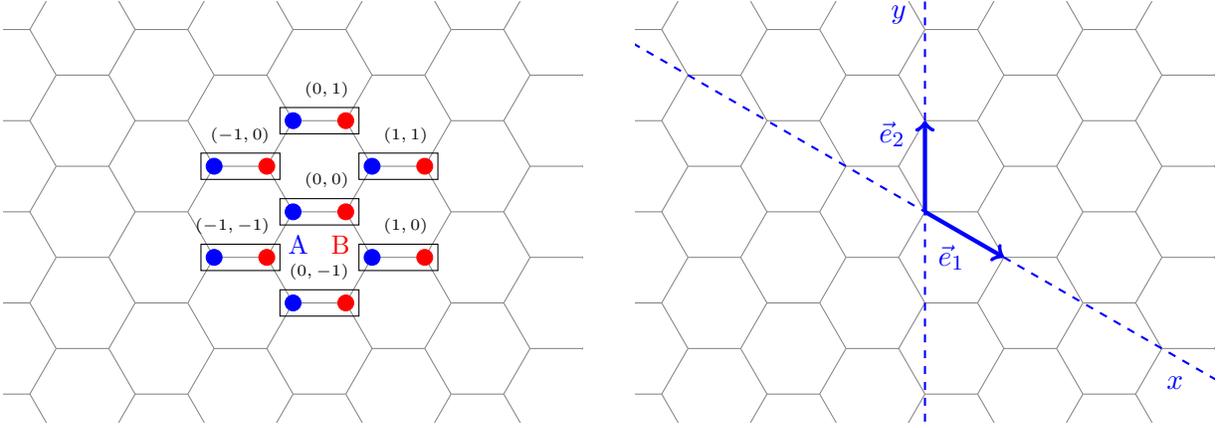
\begin{figure}[h]
\centering
\begin{tikzpicture}[scale=0.7]

\begin{scope}[xshift=0cm]
\clip (-5.5,-4) rectangle (5.5,4);
\foreach \i in {-5,-4,...,5} \foreach \j in {-5,-4,...,5} {
\foreach \a in {0,120,-120} \draw[help lines] (3*\i,2*sin{60}*\j)--+(\a:1);
\foreach \a in {0,120,-120} \draw[help lines] (3*\i+3*cos{60},2*sin{60}*\j+sin{60})--+(\a:1);}
\filldraw[blue] (0,0) circle (0.15cm);
\filldraw[red] (1,0) circle (0.15cm);
\draw (-0.25,-0.25) rectangle (1.25,0.25) node[above left] {\tiny $(0,0)$};
\node[blue] at (0.1,-0.625) {\small $\mathrm{A}$};
\node[red] at (0.9,-0.625) {\small $\mathrm{B}$};
\filldraw[blue] (0,{sqrt(3)}) circle (0.15cm);
\filldraw[red] (1,{sqrt(3)}) circle (0.15cm);
\draw (-0.25,{-0.25+sqrt(3)}) rectangle (1.25,{0.25+sqrt(3)}) node[above left] {\tiny $(0,1)$};
\filldraw[blue] (0,{-sqrt(3)}) circle (0.15cm);
\filldraw[red] (1,{-sqrt(3)}) circle (0.15cm);
\draw (-0.25,{-0.25-sqrt(3)}) rectangle (1.25,{0.25-sqrt(3)}) node[above left] {\tiny $(0,-1)$};
\filldraw[blue] (30:{sqrt(3)} ) circle (0.15cm);
\filldraw[red] (1,0)++(30:{sqrt(3)} ) circle (0.15cm);
\draw ($(30:{sqrt(3)} )+(-0.25,-0.25)$) rectangle ($(30:{sqrt(3)} )+(1.25,0.25)$) node[above left] {\tiny $(1,1)$};
\filldraw[blue] (-30:{sqrt(3)} ) circle (0.15cm);
\filldraw[red] (1,0)++(-30:{sqrt(3)} ) circle (0.15cm);
\draw ($(-30:{sqrt(3)} )+(-0.25,-0.25)$) rectangle ($(-30:{sqrt(3)} )+(1.25,0.25)$) node[above left] {\tiny $(1,0)$};
\filldraw[blue] (150:{sqrt(3)} ) circle (0.15cm);
\filldraw[red] (120:1) circle (0.15cm);
\draw ($(150:{sqrt(3)} )+(-0.25,-0.25)$) rectangle ($(150:{sqrt(3)} )+(1.25,0.25)$) node[above left] {\tiny $(-1,0)$};
\filldraw[blue] (210:{sqrt(3)} ) circle (0.15cm);
\filldraw[red] (240:1) circle (0.15cm);
\draw ($(210:{sqrt(3)} )+(-0.25,-0.25)$) rectangle ($(210:{sqrt(3)} )+(1.25,0.25)$) node[above left] {\tiny $(-1,-1)$};
\end{scope}

\begin{scope}[xshift=12cm,blue]
\clip (-5.5,-4) rectangle (5.5,4);
\foreach \i in {-5,-4,...,5} \foreach \j in {-5,-4,...,5} {
\foreach \a in {0,120,-120} \draw[help lines] (3*\i,2*sin{60}*\j)--+(\a:1);
\foreach \a in {0,120,-120} \draw[help lines] (3*\i+3*cos{60},2*sin{60}*\j+sin{60})--+(\a:1);}
\draw[thick,dashed] (-30:7)--(150:7);
\draw[thick,dashed] (0,-7)--(0,7);
\draw[ultra thick,->] (0,0)--(-30:{sqrt(3)} );
\draw[ultra thick,->] (0,0)--(0,{sqrt(3)});
\node at (4.75,-3.25) {$x$};
\node at (-0.5,3.75) {$y$};
\node at (0.5,{-sqrt(3)/2}) {$\vec{e}_1$};
\node at (-0.625,{sqrt(3)-0.25}) {$\vec{e}_2$};
\end{scope}

\end{tikzpicture}
\caption{Left: unit cells marked by rectangles on the hexagonal lattice. Each of them contains two types of vertices, $A$ and $B$ (resp. in black and in white). Right: coordinate system on the hexagonal lattice, where the positions of $A$ vertices are expressed in terms of the unit vectors $\vec{e}_1,\vec{e}_2$.}
\label{hex_coord}
\end{figure}

The second regular graph we consider is the hexagonal (or honeycomb) lattice $\mathcal{L}_{\text{H}}$. In contrast to the triangular case, there are two types of vertices on this lattice, which we call A and B (see Fig.~\ref{hex_coord}). Each vertex of type $A$ has three neighbors of type $B$, and vice versa. We choose the origin of the lattice to be of type $A$, and pick a coordinate system $\vec{r}=x\,\vec{e}_1+y\,\vec{e}_2$ where the position of each unit cell is specified by the coordinates $(x,y)\in\Z^2$ of its $A$ vertex. Each individual vertex of $\mathcal{L}_{\text{H}}$ is therefore referred to by the complete set of coordinates $(x,y;\alpha)$, with $\alpha=A,B$. Alternatively, we can use polar coordinates $(r,\varphi;\alpha)$ with the angle $\varphi$ measured counterclockwise from the $x$ axis, which are related to $(x,y;\alpha)$ through $x=r\cos\varphi+\frac{r}{\sqrt{3}}\sin\varphi$, $y=\frac{2r}{\sqrt{3}}\sin\varphi$.

\subsection{On the plane}
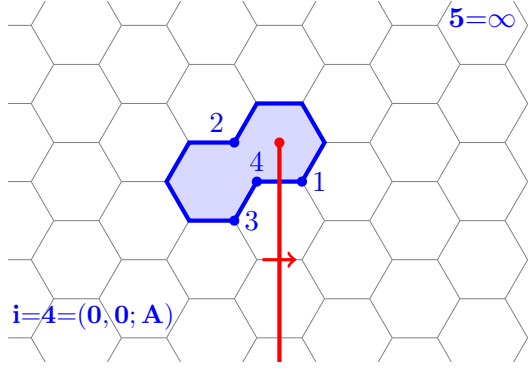
\begin{figure}[h]
\centering
\begin{tikzpicture}[scale=0.6]
\clip (-5.5,-4) rectangle (6,4);
\foreach \i in {-5,-4,...,5} \foreach \j in {-5,-4,...,5} {
\foreach \a in {0,120,-120} \draw[help lines] (3*\i,2*sin{60}*\j)--+(\a:1);
\foreach \a in {0,120,-120} \draw[help lines] (3*\i+3*cos{60},2*sin{60}*\j+sin{60})--+(\a:1);}
\filldraw[blue!15!white] (0,0)--(0:1)--++(60:1)--++(120:1)--++(180:1)--++(240:1)--++(180:1)--++(240:1)--++(300:1)--++(0:1)--++(60:1);
\draw[ultra thick,blue] (0,0)--(0:1)--++(60:1)--++(120:1)--++(180:1)--++(240:1)--++(180:1)--++(240:1)--++(300:1)--++(0:1)--++(60:1);
\filldraw[blue] (0:1) circle (0.1cm) node[right] {${1}$};
\filldraw[blue] (120:1) circle (0.1cm) node[above left] {${2}$};
\filldraw[blue] (240:1) circle (0.1cm) node[right] {${3}$};
\filldraw[blue] (0,0) circle (0.1cm) node[above] {${4}$};
\filldraw[red] (0.5,{sqrt(3)/2}) circle (0.1cm);
\draw[ultra thick,red] (0.5,{sqrt(3)/2})--(0.5,-5);
\draw[very thick,->,red] (0.125,{-sqrt(3)})--(0.875,{-sqrt(3)});
\node[blue] at (-3.625,-3.) {\small $\mathbf{i{=}4{=}(0,0;A)}$};
\node[blue] at (5,3.625) {$\mathbf{{5}{=}\infty}$};
\end{tikzpicture}
\caption{The modified graph $\xbar{\G}$ obtained by cutting the edge between nodes 2 and 4. Node 5 corresponds to the sink/root, and will eventually be sent to infinity in sandpile computations. The zipper extends down to infinity.}
\label{hex_cuts}
\end{figure}

The standard graph Laplacian on $\G=\mathcal{L}_{\text{H}}$ is defined by
\begin{equation}
\Delta_{u,v}=\begin{cases}
3&\quad\textrm{if $u=v$,}\\
-1&\quad\textrm{if $u$ and $v$ are neighbors,}\\
0&\quad\textrm{otherwise.}
\end{cases}
\end{equation}
A more appropriate way of writing $\Delta$ to compute its inverse $G$ is obtained by decomposing the lattice into unit cells \cite{SW00}. Let $a(\vec{r}_1,\vec{r}_2)\equiv a(\vec{r}_2-\vec{r}_1)$ be the $2\times 2$ adjacency matrix for the vertices of the unit cells located at $\vec{r}_1$ and $\vec{r}_2$, that is,
\begin{equation}
a_{\alpha_1,\alpha_2}(\vec{r}_2-\vec{r}_1)=\begin{cases}
1&\quad\textrm{if the vertex $\alpha_1$ of the cell $\vec{r}_1$ is a neighbor of the vertex $\alpha_2$ of the cell $\vec{r}_2$,}\\
0&\quad\textrm{otherwise.}
\end{cases}
\end{equation}
The only nonzero matrices $a(\vec{r})$ are therefore
\begin{equation}
a(0,0)=\begin{pmatrix}0 & 1\\1 & 0\end{pmatrix},\quad a(1,0)=a(1,1)=a(-1,0)^{\text{t}}=a(-1,-1)^{\text{t}}=\begin{pmatrix}0 & 0\\1 & 0\end{pmatrix}.
\end{equation}
The $2\times 2$ block entry of the Laplacian indexed by $\vec{r}_1,\vec{r}_2$ (with $\alpha_1,\alpha_2=A,B$) can then be written as follows,
\begin{equation}
\begin{split}
\Delta_{(\vec{r}_1;\alpha_1),(\vec{r}_2;\alpha_2)}&=\left\{3\,\mathbb{I}_2-a(0,0)\right\}\otimes\delta_{\vec{r}_1,\vec{r}_2}-a(1,0)\otimes\delta_{\vec{r}_1,\vec{r}_2-\vec{e}_1}-a(1,1)\otimes\delta_{\vec{r}_1,\vec{r}_2-\vec{e}_1-\vec{e}_2}\\
&\quad-a(-1,0)\otimes\delta_{\vec{r}_1,\vec{r}_2+\vec{e}_1}-a(-1,-1)\otimes\delta_{\vec{r}_1,\vec{r}_2+\vec{e}_1+\vec{e}_2}.
\end{split}
\end{equation}
Its inverse $G$ depends only on the difference $\vec{r}\equiv\vec{r}_2-\vec{r}_1=(x,y)$, and is given \cite{ADMR10} by
\begin{equation}
\begin{split}
G_{\alpha_1,\alpha_2}(x,y)&=\int_{-\pi}^{\pi}\frac{\diff\theta_1}{2\pi}\int_{-\pi}^{\pi}\frac{\diff\theta_2}{2\pi}\frac{e^{\i x\theta_1+\i y\theta_2}}{6-2\cos\theta_1-2\cos\theta_2-2\cos(\theta_1+\theta_2)}\\
&\qquad\times\begin{pmatrix}3 & 1+e^{\i\theta_1}+e^{\i(\theta_1+\theta_2)}\\1+e^{-\i\theta_1}+e^{-\i(\theta_1+\theta_2)} & 3\end{pmatrix}.
\end{split}
\label{hex_gf}
\end{equation}
It readily follows that $G_{AA}(x,y)=G_{BB}(x,y)$, $G_{AB}(-x,-y)=G_{BA}(x,y)$ and
\begin{align}
G_{AB}(x,y)&=\frac{1}{3}\Big\{G_{AA}(x,y)+G_{AA}(x+1,y)+G_{AA}(x+1,y+1)\Big\},\\
G_{BA}(x,y)&=\frac{1}{3}\Big\{G_{AA}(x,y)+G_{AA}(x-1,y)+G_{AA}(x-1,y-1)\Big\}.
\label{hex_gf_AB}
\end{align}
Moreover, we observe that $G_{AA}$ is directly related to the Green function on the triangular lattice \eqref{tri_gf}:
\begin{equation}
G^{\text{H}}_{AA}(x,y)=3\,G^{\text{T}}(x,y),
\label{Ghex_Gtri}
\end{equation}
although the variables $(x,y)$ on both sides refer to different coordinate systems (this is expected since the sublattice of $A$ vertices is triangular). For one-site sandpile probabilities on the plane, we choose the origin $i=(0,0;A)$ and its three neighbors $(0,0;B)$, $(-1,0;B)$, $(-1,-1;B)$ as nodes, and a vertical zipper anchored at the face whose lower left corner is the origin (see Fig.~\ref{hex_cuts}). In order to evaluate spanning tree probabilities, we define the graph $\xbar{\G}$ by removing the edge $((0,0;A),(-1,0;B))$ from $\G$, so that nodes 1 to 4 lie around a single face in counterclockwise order.


Since every site on the hexagonal lattice has three neighbors, the heights $h_i$ take values in $\{1,2,3\}$ for any recurrent sandpile configuration. The height-one probability is the easiest to compute, since it can be expressed in terms of the standard Green function only \cite{ADMR10}. Higher-height probabilities are given as linear combinations of spanning tree probabilities through Eq.\eqref{P_X}, which we evaluate using the same technique as for the triangular lattice. The explicit values of $\P_a(i)$ on $\G=\mathcal{L}_{\text{H}}$ are all rational numbers, equal to
\begin{equation}
\P_1(i)=\frac{1}{12}\simeq 0.083,\quad\P_2(i)=\frac{7}{24}\simeq 0.292,\quad\P_3(i)=\frac{5}{8}\simeq 0.625.
\label{hex_prob}
\end{equation}


\subsection{On upper half-planes}
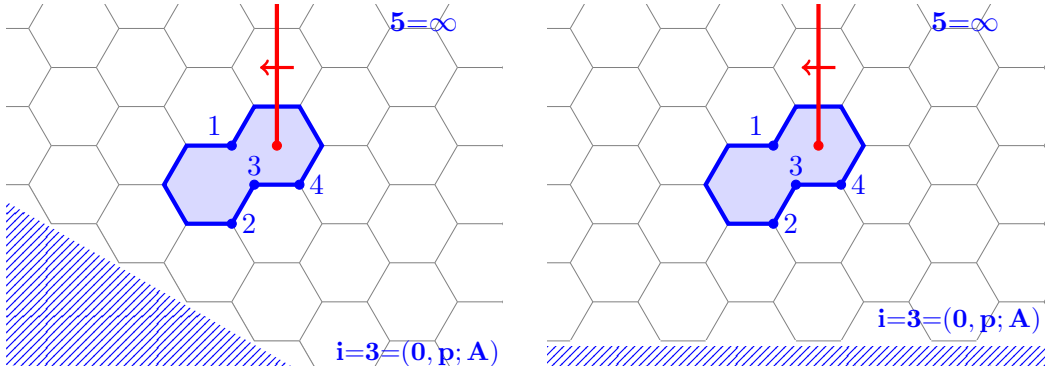
\begin{figure}[b]
\centering
\begin{tikzpicture}[scale=0.6]

\begin{scope}[xshift=0cm]
\clip (-5.5,-4) rectangle (5.5,4);
\foreach \i in {-5,-4,...,5} \foreach \j in {-5,-4,...,5} {
\foreach \a in {0,120,-120} \draw[help lines] (3*\i,2*sin{60}*\j)--+(\a:1);
\foreach \a in {0,120,-120} \draw[help lines] (3*\i+3*cos{60},2*sin{60}*\j+sin{60})--+(\a:1);}
\filldraw[blue!15!white] (0,0)--(0:1)--++(60:1)--++(120:1)--++(180:1)--++(240:1)--++(180:1)--++(240:1)--++(300:1)--++(0:1)--++(60:1);
\draw[ultra thick,blue] (0,0)--(0:1)--++(60:1)--++(120:1)--++(180:1)--++(240:1)--++(180:1)--++(240:1)--++(300:1)--++(0:1)--++(60:1);
\filldraw[blue] (0:1) circle (0.1cm) node[right] {${4}$};
\filldraw[blue] (120:1) circle (0.1cm) node[above left] {${1}$};
\filldraw[blue] (240:1) circle (0.1cm) node[right] {${2}$};
\filldraw[blue] (0,0) circle (0.1cm) node[above] {${3}$};
\filldraw[red] (0.5,{sqrt(3)/2}) circle (0.1cm);
\draw[ultra thick,red] (0.5,{sqrt(3)/2})--(0.5,5);
\draw[very thick,->,red] (0.875,{3*sqrt(3)/2})--(0.125,{3*sqrt(3)/2});
\filldraw[white] (-6,0)--++(0:1)--++(300:1)--++(0:1)--++(300:1)--++(0:1)--++(300:1)--++(0:1)--++(300:1)--++(0:1)--++(300:1)--(-6,-5)--(-6,0);
\draw[help lines] (-6,0)--++(0:1)--++(300:1)--++(0:1)--++(300:1)--++(0:1)--++(300:1)--++(0:1)--++(300:1)--++(0:1)--++(300:1);
\fill[pattern=north east lines,pattern color=blue,shift={(-5.5,-4)}] (0,0)--(0,3.625)--++(330:8)--(0,0);
\node[blue] at (3.625,-3.75) {\small $\mathbf{i{=}3{=}(0,p;A)}$};
\node[blue] at (3.75,3.625) {$\mathbf{{5}{=}\infty}$};
\end{scope}

\begin{scope}[xshift=12cm]
\clip (-5.5,-4) rectangle (5.5,4);
\foreach \i in {-5,-4,...,5} \foreach \j in {-5,-4,...,5} {
\foreach \a in {0,120,-120} \draw[help lines] (3*\i,2*sin{60}*\j)--+(\a:1);
\foreach \a in {0,120,-120} \draw[help lines] (3*\i+3*cos{60},2*sin{60}*\j+sin{60})--+(\a:1);}
\filldraw[blue!15!white] (0,0)--(0:1)--++(60:1)--++(120:1)--++(180:1)--++(240:1)--++(180:1)--++(240:1)--++(300:1)--++(0:1)--++(60:1);
\draw[ultra thick,blue] (0,0)--(0:1)--++(60:1)--++(120:1)--++(180:1)--++(240:1)--++(180:1)--++(240:1)--++(300:1)--++(0:1)--++(60:1);
\filldraw[blue] (0:1) circle (0.1cm) node[right] {${4}$};
\filldraw[blue] (120:1) circle (0.1cm) node[above left] {${1}$};
\filldraw[blue] (240:1) circle (0.1cm) node[right] {${2}$};
\filldraw[blue] (0,0) circle (0.1cm) node[above] {${3}$};
\filldraw[red] (0.5,{sqrt(3)/2}) circle (0.1cm);
\draw[ultra thick,red] (0.5,{sqrt(3)/2})--(0.5,5);
\draw[very thick,->,red] (0.875,{3*sqrt(3)/2})--(0.125,{3*sqrt(3)/2});
\node[blue] at (3.625,-3) {\small $\mathbf{i{=}3{=}(0,p;A)}$};
\node[blue] at (3.75,3.625) {$\mathbf{{5}{=}\infty}$};
\filldraw[white] (-5.5,{-2*sqrt(3)}) rectangle (5.5,-4);
\foreach \x in {-2,-1,0,1}{\draw[help lines] ({3*\x},{-2*sqrt(3)})--({3*\x+1},{-2*sqrt(3)});}
\fill[pattern=north east lines,pattern color=blue] (-5.5,{-2*sqrt(3)-0.125}) rectangle (5.5,-4);
\end{scope}

\end{tikzpicture}
\caption{The modified graph $\xbar{\G}$ obtained by cutting the edge between nodes 1 and 3 on half-lattices with a principal boundary (on the left) and a horizontal boundary (on the right). The sink corresponds to node 5 and is sent to infinity. The zipper extends up to infinity.}
\label{hex_uhp_cuts}
\end{figure}

In addition to the lattice $\mathcal{L}_{\text{H}}$, we consider half-lattices with two kinds of boundaries:  one parallel to the $x$ axis, and a horizontal one (see Fig.~\ref{hex_uhp_cuts}). For both half-planes, which, following \cite{ADMR10}, we refer respectively to as \emph{principal} and \emph{horizontal}, we choose the reference site $i=(0,p;A)$ with $p\gg 1$. As for the full lattice, we select $i$, its three neighbors and the sink as nodes on a modified graph, here obtained by cutting the edge between $i$ and its neighbor $(-1,p;B)$. For simplicity, the zipper is taken as a path on the dual graph starting on a face adjacent to $i$ and extending up to infinity. The edges with a nontrivial parallel transport $z\in\C^*$ are the following: $((0,k;B),(0,k;A))$ for $k\ge p+1$. The same arguments as in Section \ref{sec3} are used to write the fractions of spanning trees $X_q(i)$ in terms of the Green function $G$ of the half-lattice of interest and the Green function derivative $G'$ associated with the zipper.

Let us first look at the principal half-plane, whose boundary consists in vertices of the form $(x,y{=}1;A)$, as depicted in the left panel of Fig.~\ref{hex_uhp_cuts}. The corresponding Green function can be written in terms of the full-plane Green function, for either closed or open boundary conditions, using the image method. For the closed boundary, each site is mirrored through the reflection axis $y=2/3$, so the Green function reads \cite{ADMR10}
\begin{equation}
\begin{split}
&G^{\mathrm{cl}}_{(x_1,y_1;\alpha_1),(x_2,y_2;A)}=G_{(x_1,y_1;\alpha_1),(x_2,y_2;A)}+G_{(x_1,y_1;\alpha_1),(x_2-y_2,1-y_2;B)},\\
&G^{\mathrm{cl}}_{(x_1,y_1;\alpha_1),(x_2,y_2;B)}=G_{(x_1,y_1;\alpha_1),(x_2,y_2;B)}+G_{(x_1,y_1;\alpha_1),(x_2-y_2+1,1-y_2;A)}.
\end{split}
\label{hex_gf_princ_cl}
\end{equation}
Proceeding as in \eqref{tri_gp_op}, we may write the Green function derivative with respect to the zipper represented in Fig.~\ref{hex_cuts} in terms of that on the full lattice. For instance if $\alpha_1=\alpha_2=A$, we find the relation
\begin{equation}
\begin{split}
G^{\prime\,\mathrm{cl}}_{(x_1,y_1;A),(x_2,y_2;A)}&=G'_{(-x_1,p+1-y_1;B),(-x_2,p+1-y_2;B)}-G'_{(-x_1,p+1-y_1;B),(y_2-x_2,p+y_2;A)}\\
&\quad-G'_{(y_1-x_1,p+y_1;A),(-x_2,p+1-y_2;B)}+G'_{(y_1-x_1,p+y_1;A),(y_2-x_2,p+y_2;A)}.
\end{split}
\end{equation}

The open case is subtler, as the mirror image of a vertex of type $A$ with respect to the natural reflection line $y=1/3$ does not belong to the lattice. However, one can use the Poisson equation to get a suitable expression for the Green function \cite{ADMR10}:
\begin{equation}
\begin{split}
G^{\mathrm{op}}_{(x_1,y_1;\alpha_1),(x_2,y_2;A)}&=G_{(x_1,y_1;\alpha_1),(x_2,y_2;A)}-\frac{1}{3}\Big[G_{(x_1,y_1;\alpha_1),(x_2-y_2,-y_2;B)}\\
&\qquad+G_{(x_1,y_1;\alpha_1),(x_2-y_2-1,-y_2;B)}+G_{(x_1,y_1;\alpha_1),(x_2-y_2,1-y_2;B)}\Big],\\
G^{\mathrm{op}}_{(x_1,y_1;\alpha_1),(x_2,y_2;B)}&=G_{(x_1,y_1;\alpha_1),(x_2,y_2;B)}-G_{(x_1,y_1;\alpha_1),(x_2-y_2,-y_2;B)}.
\end{split}
\label{hex_gf_princ_op}
\end{equation}
Here the Green function derivatives on the half-lattice and on the full lattice are connected through
\begin{equation}
\begin{split}
G^{\prime\,\mathrm{op}}_{(x_1,y_1;B),(x_2,y_2;B)}&=G'_{(-x_1,p+1-y_1;A),(-x_2,p+1-y_2;A)}+G'_{(-x_1,p+1-y_1;A),(y_2-x_2,p+1+y_2;A)}\\
&\quad+G'_{(y_1-x_1,p+1+y_1;A),(-x_2,p+1-y_2;A)}+G'_{(y_1-x_1,p+1+y_1;A),(y_2-x_2,p+1+y_2;A)}
\end{split}
\end{equation}
for $\alpha_1=\alpha_2=B$. Similar relations hold for other values of $\alpha_1,\alpha_2$.

We use these formulas together with the integral representation of the Green function on the full lattice \eqref{hex_gf} to compute $G^{\mathrm{cl,op}}$ and $G^{\prime\,\mathrm{cl,op}}$ as series expansions in $1/p$ (this is discussed in the appendix for the triangular lattice; the treatment of the hexagonal lattice is very similar). We find that one-site probabilities on the principal half-plane take the form
\begin{equation}
\sigma^{\text{princ}}_a(r) \equiv \P^{\text{princ}}_a(r)-\P_a = \frac{1}{r^2}\left(c^{\text{princ}}_a+d^{\text{princ}}_a\log r\right)+\mathcal{O}(r^{-3}\log r),
\label{hex_princ_uhp_sig}
\end{equation}
where $\P_a$ denotes the one-site probability on the full lattice, and $r$ is the Euclidean distance between the reference site $i=(0,p;A)$ and the reflection axis used in the image method, namely
\begin{equation}
r=\frac{\sqrt{3}}{2}p-\frac{1}{\sqrt{3}}\quad\textrm{(closed b.c.)}, \qquad
r=\frac{\sqrt{3}}{2}p-\frac{1}{2\sqrt{3}}\quad\textrm{(open b.c.)}.
\end{equation}
The numerical values of the coefficients $c^{\text{princ}}_a,d^{\text{princ}}_a$ for both types of boundary conditions are collected in Table~\ref{hex_princ_uhp_coef}. We note the distinctive change of sign between the two boundary conditions in the most dominant terms (rational for the height 1, logarithmic for the higher heights), as encountered on the square half-lattice \cite{JPR06}.

\begin{table}[h]
\centering
\renewcommand{\arraystretch}{1.8}
\tabcolsep11pt
\begin{tabular}{|c|cc|cc|}
\hline
 & $c^{\text{princ},\mathrm{cl}}_a$ & $d^{\text{princ},\mathrm{cl}}_a$ & $c^{\text{princ},\mathrm{op}}_a$ & $d^{\text{princ},\mathrm{op}}_a$\\
\hline
$a=1$ & $-\frac{1}{16\sqrt{3}\pi}$ & $0$ & $\frac{1}{16\sqrt{3}\pi}$ & $0$\\
$a=2$ & $-\frac{3}{16\pi^2}\left(\gamma{+}\frac{1}{2}\text{{\small $\log 48$}}\right){+}\frac{15}{64\pi^2}$\hspace{-0.5cm} & $-\frac{3}{16\pi^2}$ & $\frac{3}{16\pi^2}\left(\gamma{+}\frac{1}{2}\text{{\small $\log 48$}}\right){-}\frac{9}{64\pi^2}$\hspace{-0.5cm} & $\frac{3}{16\pi^2}$\\
$a=3$ & $\frac{3}{16\pi^2}\left(\gamma{+}\frac{1}{2}\text{{\small $\log 48$}}\right){-}\frac{45{-}4\sqrt{3}\pi}{192\pi^2}$\hspace{-0.5cm} & $\frac{3}{16\pi^2}$ & $-\frac{3}{16\pi^2}\left(\gamma{+}\frac{1}{2}\text{{\small $\log 48$}}\right){+}\frac{27{-}4\sqrt{3}\pi}{192\pi^2}$\hspace{-0.5cm} & $-\frac{3}{16\pi^2}$\\
\hline
\end{tabular}
\caption{Coefficients for one-site probabilities on the hexagonal half-lattice with a principal boundary.}
\label{hex_princ_uhp_coef}
\end{table}

Let us now turn to the horizontal half-plane (drawn on the right panel of Fig.~\ref{hex_uhp_cuts}), whose boundary sites $(x,y;\alpha)$ satisfy the equality $x=2y-2$. For an open boundary, the image method allows one to write the Green function as \cite{ADMR10}
\begin{equation}
G^{\mathrm{op}}_{(x_1,y_1;\alpha_1),(x_2,y_2;\alpha_2)}=G_{(x_1,y_1;\alpha_1),(x_2,y_2;\alpha_2)}-G_{(x_1,y_1;\alpha_1),(x_2,x_2-y_2+1;\alpha_2)}.
\label{hex_gf_hor_op}
\end{equation}
With respect to the zipper pictured in Fig.~\ref{hex_uhp_cuts}, the Green function derivative reads:
\begin{equation}
\begin{split}
G^{\prime\,\text{op}}_{(x_1,y_1;\alpha_1),(x_2,y_2;\alpha_2)}&=G'_{(-x_1,p+1-y_1;\bar{\alpha}_1),(-x_2,p+1-y_2;\bar{\alpha}_2)}-G'_{(-x_1,p+1-y_1;\bar{\alpha}_1),(-x_2,p-x_2+y_2;\bar{\alpha}_2)}\\
&\quad -G'_{(-x_1,p-x_1+y_1;\bar{\alpha}_1),(-x_2,p+1-y_2;\bar{\alpha}_2)}+G'_{(-x_1,p-x_1+y_1;\bar{\alpha}_1),(-x_2,p-x_2+y_2;\bar{\alpha}_2)},
\end{split}
\end{equation}
where $G'$ is the derivative on the full lattice with a zipper going down to infinity (see Fig.~\ref{hex_cuts}), and $\bar{\alpha}=B$ (resp. $A$) if $\alpha=A$ (resp. $B$). For a closed boundary, we have not been able to find a suitable reflection axis for the image method, so we only give results for the open half-plane. 

Let us define $r$ as the distance between the reference site $i=(0,p;A)$ and the reflection axis used in the image method $y=(x+1)/2$, i.e. $r=p-1/2$. Up to third-order terms, one-site probabilities on the open horizontal half-plane are given by
\begin{equation}
\sigma^{\text{hor,op}}_a(r) \equiv \P^{\text{hor,op}}_a(r)-\P_a = \frac{1}{r^2}\left(c^{\text{hor,op}}_a+d^{\text{hor,op}}_a\log r\right)+\mathcal{O}(r^{-3}\log r),
\label{hex_hor_uhp_sig}
\end{equation}
with coefficients $c^{\text{hor,op}}_a,d^{\text{hor,op}}_a$ given in Table \ref{hex_hor_uhp_coef}. As expected, one-site probabilities on the principal and horizontal half-planes with open boundary conditions coincide at order $1/r^2$, namely $c^{\text{hor,op}} = c^{\text{princ,op}}$ and $d^{\text{hor,op}} = d^{\text{princ,op}}$.

\begin{table}[htbp]
\centering
\large
\renewcommand{\arraystretch}{1.8}
\tabcolsep12pt
\begin{tabular}{|c|cc|}
\hline
 & $c^{\text{hor},\mathrm{op}}_a$ & $d^{\text{hor},\mathrm{op}}_a$\\
\hline
$a=1$ & $\frac{1}{16\sqrt{3}\pi}$ & $0$\\
$a=2$ & $\frac{3}{16\pi^2}\left(\gamma+\frac{1}{2}\text{{\small $\log 48$}}\right)-\frac{9}{64\pi^2}$ & $\frac{3}{16\pi^2}$\\
$a=3$ & $-\frac{3}{16\pi^2}\left(\gamma+\frac{1}{2}\text{{\small $\log 48$}}\right)+\frac{27-4\sqrt{3}\pi}{192\pi^2}$ & $-\frac{3}{16\pi^2}$\\
\hline
\end{tabular}
\caption{Coefficients for one-site probabilities on the hexagonal half-lattice with a horizontal open boundary.}
\label{hex_hor_uhp_coef}
\end{table}


\subsection{On boundaries}
In this subsection, we consider probabilities involving vertices located on the boundary of the two hexagonal upper half-planes described in the previous subsection. Due to the small number of admissible heights (two on a closed boundary, three on an open one), the calculations are quite straightforward, as they do not require the use of a nontrivial connection on the graph. Indeed, as explained in Section~\ref{sec3}, height-one probabilities can be computed in terms of the standard Green function only. The same is true for height-three probabilities on an open boundary, see below, yielding height-two probabilities by subtraction.

We shall compute one-site and two-site probabilities on the boundary of the two half-lattices described above. The simplest case is the principal boundary with closed boundary conditions, since the height of a closed boundary site takes on the values 1 or 2. Therefore height-two probabilities can be obtained by subtraction from height-one probabilities, which can be computed via defect matrices. This has been done in \cite{ADMR10} with the following results,
\begin{align}
&\P^{\text{princ},\text{cl}}_1 = \frac{\sqrt{3}}{\pi}-\frac{1}{3},\quad\P^{\text{princ},\text{cl}}_2 = \frac{4}{3}-\frac{\sqrt{3}}{\pi},\\
&\P^{\text{princ},\text{cl}}_{1,1}(i,j) - \P^{\text{princ},\text{cl}}_1\,\P^{\text{princ},\text{cl}}_1 = -\frac{3}{16\pi^2 x^4}+\ldots,
\end{align}
where $x=|x_2-x_1|$ denotes the Euclidean distance between sites $i=(x_1,1;A)$ and $j=(x_2,1;A)$.

For open boundary conditions on the principal half-plane, boundary heights $h_i$ take their value in $\{1,2,3\}$. The height-three probability at site $i$ can be evaluated as follows \cite{PR05a}: define a new Laplacian $\widetilde{\Delta}$ such that $\widetilde{\Delta}_{i,i}=\Delta_{i,i}{-}1$, with $\widetilde{\Delta}$ and $\Delta$ coinciding everywhere else. The burning algorithm \cite{MD92} gives a bijection between recurrent configurations with $h_i=3$ and spanning trees that use the edge between $i$ and the sink $s$. As $\det\widetilde{\Delta}$ counts precisely the number of spanning trees that do not use that particular edge, it follows that
\begin{equation}
\P_3(i)=1-\frac{\det\widetilde{\Delta}}{\det\Delta}=\left(\Delta^{-1}\right)_{i,i}.
\label{hex_P3b}
\end{equation}
The remaining height probability $\P_2(i)$ can be obtained from the relation $\sum_{a=1}^{3}\P_a(i)=1$. With the appropriate Green function \eqref{hex_gf_princ_op}, we find the following one-site and two-site probabilities for the open boundary conditions\footnote{Only height-one probabilities were given in \cite{ADMR10}.}:
\begin{align}
&\P^{\text{princ},\text{op}}_1 = \frac{11}{36}+\frac{4}{\sqrt{3}\pi}-\frac{9}{\pi^2},\quad\P^{\text{princ},\text{op}}_2 = -\frac{7}{36}-\frac{2}{\sqrt{3}\pi}+\frac{9}{\pi^2},\quad\P^{\text{princ},\text{op}}_3 = \frac{8}{9}-\frac{2}{\sqrt{3}\pi},\\
&\P^{\text{princ},\text{op}}_{a,b}(i,j) - \P^{\text{princ},\text{op}}_a\,\P^{\text{princ},\text{op}}_b = -\frac{\alpha^{\text{princ,op}}_a\alpha^{\text{princ,op}}_b}{4x^4}+\ldots, \qquad a,b=1,2,3,
\label{hex_princ_bd_2prob}
\end{align}
with $\alpha^{\text{princ,op}}_1=\frac{11}{2\sqrt{3}\pi}-\frac{9}{\pi^2}$, $\alpha^{\text{princ,op}}_2=-\frac{7}{2\sqrt{3}\pi}+\frac{9}{\pi^2}$, $\alpha^{\text{princ,op}}_3=-\frac{2}{\sqrt{3}\pi}$ (up to a global sign).

On the horizontal half-plane with open boundary conditions, we obtain similar results using the Green function \eqref{hex_gf_hor_op}:
\begin{align}
&\P^{\text{hor},\text{op}}_1(i)=-\frac{37}{36}+\frac{8}{\sqrt{3}\pi}-\frac{3}{\pi^2},\quad\P^{\text{hor},\text{op}}_2(i)=\frac{55}{36}-\frac{8}{\sqrt{3}\pi}+\frac{3}{\pi^2},\quad\P^{\text{hor},\text{op}}_3(i)=\frac{1}{2},\\
&\P^{\text{hor},\text{op}}_{a,b}(i,j) - \P^{\text{hor},\text{op}}_a\,\P^{\text{hor},\text{op}}_b = -\frac{\alpha^{\text{hor,op}}_a\alpha^{\text{hor,op}}_b}{4x^4}+\ldots, \qquad a,b=1,2,3,
\label{hex_hor_bd_2prob}
\end{align}
with $\alpha^{\text{hor,op}}_1=-\frac{1}{3\sqrt{3}\pi}+\frac{1}{\pi^2}$, $\alpha^{\text{hor,op}}_2=\frac{5}{6\sqrt{3}\pi}-\frac{1}{\pi^2}$, $\alpha^{\text{hor,op}}_3=-\frac{1}{2\sqrt{3}\pi}$ (again up to a global sign). 

More generally, multisite boundary probabilities on hexagonal half-lattices could easily be computed using Eqs. \eqref{X0_comp} and \eqref{hex_P3b}. For the triangular half-plane with an open boundary, the situation is more complicated, as there are three nontrivial probabilities to evaluate separately, namely 2, 3 and 4 (heights 5 and 6 can be handled in the same way as height 3 on the hexagonal lattice).


\section{Discussion}
\label{sec6}

On the square lattice, joint height probabilities at sites separated by large distances have been considered on the full plane and in the bulk of the upper half-plane. The probabilities involving only unit heights are given by rational functions of the distances \cite{MD91,BIP93,MR01}. In contrast, probabilities including at least one height strictly larger than one contain logarithmic terms \cite{JPR06,PGPR10}. In light of these results, the variables $h_a(i)\equiv\delta_{h(i),a}-\P_a$ in the bulk of the lattice (far from the boundaries) are expected to converge to conformal fields $h_a(z,\bar{z})$ of weight $(1{,}1)$ in the scaling limit, where $\P_a$ is the one-site probability on the full infinite lattice, $a=1,2,3,4$. The field $h_1(z,\bar{z})$ has been identified as a primary field $\phi$, degenerate at level two, and the other fields $h_{a>1}(z,\bar{z})$ as  (appropriately normalized) logarithmic partners $\psi$ of $\phi$, in a rank-2 Jordan cell \cite{PR05b,JPR06}. More precisely, if $\psi$ refers to a fixed logarithmic partner of $\phi$, the height fields have been expressed as linear combinations $h_a(z,\bar{z})=\alpha_a\psi(z,\bar{z})+\beta_a\phi(z,\bar{z})$, with $\alpha_1=0$. Our results for the subtracted one-point probabilities $\sigma_a(r)$ on the triangular and hexagonal half-lattices, given respectively by Eqs. \eqref{tri_uhp_sig}, \eqref{hex_princ_uhp_sig} and \eqref{hex_hor_uhp_sig} are perfectly consistent with this picture.

The logarithmic conformal field theory in play has further been described by the identification of the operator $\mu$ implementing the change of boundary condition, from open to closed or vice versa. This chiral field has been found to be primary, of weight $-1/8$ \cite{Rue02}. It follows that the correlators $\langle h_a(z,\bar{z})\rangle_{\text{op}}$ and $\langle h_a(z,\bar{z})\rangle_{\text{cl}}$ corresponding to the scaling limit of $\sigma^{\text{op}}_a(r)$ and $\sigma^{\text{cl}}_a(r)$ must be related by the following equations \cite{JPR06}:
\begin{equation}
c^{\,\text{op}}_a=-c^{\,\text{cl}}_a-\tfrac{1}{2}d^{\,\text{cl}}_a,\quad d^{\,\text{op}}_a=-d^{\,\text{cl}}_a.
\end{equation}
Again we find these relations to be in full agreement with the coefficients in Table \ref{hex_princ_uhp_coef} for the principal hexagonal half-plane. Unfortunately, we cannot make the comparison for the other two half-lattices considered in this paper, namely the triangular half-plane and the horizontal hexagonal half-plane, as we have not been able to compute the Green function for a closed boundary.

In contrast to their bulk analogues, all subtracted probabilities between boundary sites of the square half-lattice are given by rational functions. The boundary height fields have been identified as nonlogarithmic fields, which have weight 2 \cite{Iva94,Jen05b,PR05a}. Moreover, for open boundary conditions, the observation that height correlations on the square lattice factorize signal the fact that all boundary height fields are proportional to a single conformal field (this assumption was shown to be false for a closed boundary). Equations \eqref{hex_princ_bd_2prob} and \eqref{hex_hor_bd_2prob} indeed confirm this assertion for both hexagonal half-planes with open boundary. One notices that the factorization property also holds for the \emph{closed} hexagonal half-plane; however, it is merely due to the fact the two height fields are required to sum to zero.


\subsection*{Acknowledgments}
This work is supported by the Belgian Interuniversity Attraction Poles Program P7/18 through the network DYGEST (Dynamics, Geometry and Statistical Physics). PR is Senior Research Associate of the Belgian Fonds National de la Recherche Scientifique (FNRS).


\appendix
\section{Green functions and derivatives}
\label{app}

In this appendix, we first recall some exact values of the Green function of the triangular lattice for small distances, as well as its asymptotic expansion for large distances. Then we discuss the methods we used to compute the Green function derivative on the full plane, with respect to the zipper depicted in Fig.~\ref{tri_zip}. We work out an example in details for the short-distance regime, to illustrate the general principle of computation, and provide some explanations and two typical values for the large-distance regime. Finally, for a pair of graphs $(\G,\xbar{\G})$ related to one another by the addition or removal of a few edges, we recall the relation between their respective Laplacians and Green functions (this was needed in Sections \ref{sec4} and \ref{sec5}). The analogous developments for the hexagonal lattice will be skipped (indeed, remember that the Green functions of both lattices are related through Eq.~\eqref{Ghex_Gtri}).


\subsection{Values of Green functions}
\label{app_sec1}
In the coordinate system of Section~\ref{sec4}, the Green function of the triangular lattice was given in integral form in \eqref{tri_gf},
\begin{equation}
G_{(x_1,y_1),(x_2,y_2)}=G(x_1{-}x_2,y_1{-}y_2)=\int_{-\pi}^{\pi}\frac{\diff\theta_1}{2\pi}\int_{-\pi}^{\pi}\frac{\diff\theta_2}{2\pi}\frac{e^{\i(x_1-x_2)\theta_1+\i(y_1-y_2)\theta_2}}{6-2\cos\theta_1-2\cos\theta_2-2\cos(\theta_1+\theta_2)}.
\label{tri_gfs}
\end{equation}
In Table \ref{tri_gf_table}, we list the value of the Green function for small separations \cite{AvS99}, where $G_{0,0}\equiv G(0,0)$ is the divergent part of the integral \eqref{tri_gfs}. The large-distance expansion of $G(x,y)$ can be computed from the corresponding result on the hexagonal lattice \cite{ADMR10} and is given by
\begin{equation}
G(x,y)=G_{0,0}-\frac{1}{2\sqrt{3}\pi}\left(\log r+\gamma+\frac{1}{2}\log 12\right)+\frac{\cos 6\varphi}{60\sqrt{3}\pi r^4}+\mathcal{O}(r^{-6}),
\label{tri_gf_asy}
\end{equation}
where $r=\sqrt{x^2+y^2-xy}\gg 1$ and $\varphi$ is the angle between the horizontal axis and $(x,y)$, i.e. $x=r\cos\varphi+\frac{r}{\sqrt{3}}\sin\varphi$ and $y=\frac{2r}{\sqrt{3}}\sin\varphi$. Here $\gamma= 0.577216...$ is the Euler constant.

\begin{table}[htbp]
\centering
\large
\renewcommand{\arraystretch}{1.8}
\tabcolsep12pt
\begin{tabular}{|c|ccccc|}
\hline
$G(x,y)-G_{0,0}$ & $y=-2$ & $y=-1$ & $y=0$ & $y=1$ & $y=2$\\
\hline
$x=-2$ & $-\frac{4}{3}+\frac{2\sqrt{3}}{\pi}$ & $\frac{1}{3}-\frac{\sqrt{3}}{\pi}$ & $-\frac{4}{3}+\frac{2\sqrt{3}}{\pi}$ & $\frac{5}{2}-\frac{5\sqrt{3}}{\pi}$ & $-8+\frac{14\sqrt{3}}{\pi}$\\
$x=-1$ & $\frac{1}{3}-\frac{\sqrt{3}}{\pi}$ & $-\frac{1}{6}$ & $-\frac{1}{6}$ & $\frac{1}{3}-\frac{\sqrt{3}}{\pi}$ & $\frac{5}{2}-\frac{5\sqrt{3}}{\pi}$\\
$x=0$ & $-\frac{4}{3}+\frac{2\sqrt{3}}{\pi}$ & $-\frac{1}{6}$ & 0 & $-\frac{1}{6}$ & $-\frac{4}{3}+\frac{2\sqrt{3}}{\pi}$\\
$x=1$ & $\frac{5}{2}-\frac{5\sqrt{3}}{\pi}$ & $\frac{1}{3}-\frac{\sqrt{3}}{\pi}$ & $-\frac{1}{6}$ & $-\frac{1}{6}$ & $\frac{1}{3}-\frac{\sqrt{3}}{\pi}$\\
$x=2$ & $-8+\frac{14\sqrt{3}}{\pi}$ & $\frac{5}{2}-\frac{5\sqrt{3}}{\pi}$ & $-\frac{4}{3}+\frac{2\sqrt{3}}{\pi}$ & $\frac{1}{3}-\frac{\sqrt{3}}{\pi}$ & $-\frac{4}{3}+\frac{2\sqrt{3}}{\pi}$\\
\hline
\end{tabular}
\caption{Values of the Green function of the triangular lattice for small distances.}
\label{tri_gf_table}
\end{table}


\subsection{Green function derivative for small distances}
\label{app_sec2}
The definition \eqref{gp_def} of the Green function derivative $G'$ with respect to the zipper in Fig.~\ref{tri_zip} yields:
\begin{equation}
\begin{split}
G'_{(x_1,y_1),(x_2,y_2)}&=\sum_{k=0}^{\infty}\Big[G_{(x_1,y_1),(1,-k)}G_{(0,-k),(x_2,y_2)}-G_{(x_1,y_1),(0,-k)}G_{(1,-k),(x_2,y_2)}\\
&\quad\quad+G_{(x_1,y_1),(1,-k)}G_{(0,-k-1),(x_2,y_2)}-G_{(x_1,y_1),(0,-k-1)}G_{(1,-k),(x_2,y_2)}\Big].
\end{split}
\label{tri_gp_exp}
\end{equation}
For generic vertices $u_i=(x_i,y_i)$, $G'$ must be computed through an explicit summation over all zipper edges using the integral representation of the Green function \eqref{tri_gfs}. However, if $u_1,u_2$ are close to the head of the zipper (i.e. close to the origin of the lattice), we can find the value $G'$ using the following symmetries and transformation properties \cite{KW15,PR17}:
\begin{enumerate}[(a)]
\item $G'$ is antisymmetric, $G'_{u_1,u_2} = -G'_{u_2,u_1}$;
\item $G'$ only depends on the relative position of its arguments with respect to zipper edges;
\item Deforming the zipper in the direction of its arrow on the dual lattice while keeping its endpoints fixed changes the value of $G'_{u_1,u_2}$ by $-G_{u_1,u_2}$ (resp. $+G_{u_1,u_2}$) if the zipper crosses $u_1$ (resp. $u_2$) only, and $G'_{u_1,u_2}$ remains unchanged if the zipper crosses both $u_1,u_2$ or neither of them:
\begin{equation}
G^{\prime\,\textrm{new}}_{u_1,u_2}-G^{\prime\,\textrm{old}}_{u_1,u_2}=\begin{cases}
-G_{u_1,u_2}&\quad\textrm{if the zipper crosses $u_1$ only,}\\
+G_{u_1,u_2}&\quad\textrm{if the zipper crosses $u_2$ only,}\\
0&\quad\textrm{otherwise.}
\end{cases}
\end{equation}
\end{enumerate}
The first two assertions come directly from Eq.~\eqref{gp_def} and the invariance under translations of the triangular lattice. The third one follows from the fact that $G'$ satisfies a Poisson-like equation \cite{PR17}:
\begin{equation}
\left(\Delta G'\right)_{u_1,u_2}=-\left(G'\Delta\right)_{u_2,u_1}=\sum_{(k,\l):\,\phi_{k,\l}=z}\Big[-\delta_{u_1,k}\,G_{\l,u_2}+\delta_{u_1,\l}\,G_{k,u_2}\Big].
\end{equation}

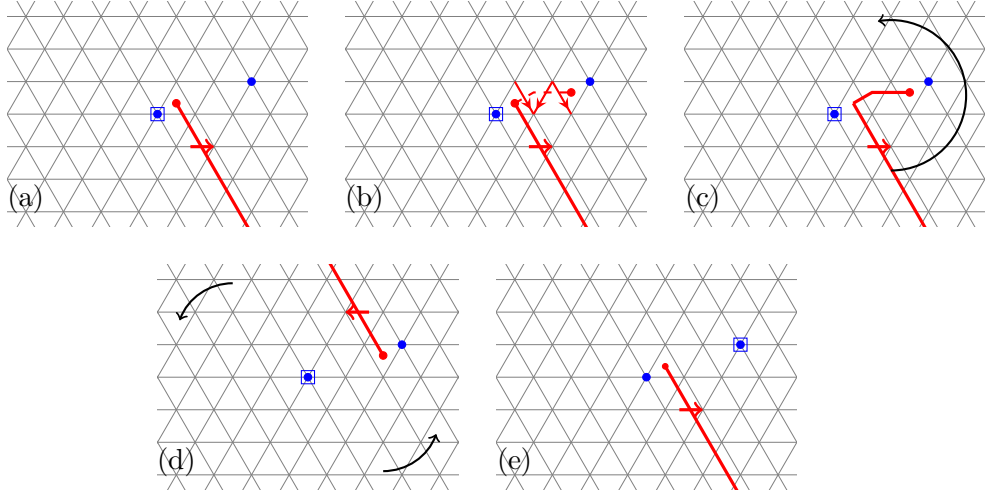
\begin{figure}[t]
\centering
\begin{tikzpicture}[scale=0.5]
\tikzstyle arrowstyle=[scale=1]
\tikzstyle directed=[postaction={decorate,decoration={markings,mark=at position 0.85 with {\arrow[arrowstyle]{stealth}}}}]

\begin{scope}[xshift=0cm]
\clip (-4,-3) rectangle (4,3);
\foreach \x in {-5,-4,...,5}{
\draw[help lines] (-5,{\x*sqrt(3)/2})--(5,{\x*sqrt(3)/2});
\draw[xshift=\x cm,help lines] (60:6)--(240:6);
\draw[xshift=\x cm,help lines] (120:6)--(300:6);
}
\filldraw[blue] (0,0) circle (0.1cm);
\draw[blue] (-0.175,-0.175) rectangle (0.175,0.175);
\filldraw[blue] (2.5,{sqrt(3)/2}) circle (0.1cm);
\filldraw[red] (0.5,{sqrt(3)/6}) circle (0.1cm);
\draw[very thick,red] ++(0.5,{sqrt(3)/6} )--++(-60:5);
\draw[very thick,->,red] (0.875,{-sqrt(3)/2})--(1.5,{-sqrt(3)/2});
\node at (-3.5,-2.25) {(a)};
\end{scope}

\begin{scope}[xshift=9cm]
\clip (-4,-3) rectangle (4,3);
\foreach \x in {-5,-4,...,5}{
\draw[help lines] (-5,{\x*sqrt(3)/2})--(5,{\x*sqrt(3)/2});
\draw[xshift=\x cm,help lines] (60:6)--(240:6);
\draw[xshift=\x cm,help lines] (120:6)--(300:6);
}
\filldraw[blue] (0,0) circle (0.1cm);
\draw[blue] (-0.175,-0.175) rectangle (0.175,0.175);
\filldraw[blue] (2.5,{sqrt(3)/2}) circle (0.1cm);
\filldraw[red] (0.5,{sqrt(3)/6}) circle (0.1cm);
\draw[very thick,red] ++(0.5,{sqrt(3)/6} )--++(-60:5);
\draw[very thick,->,red] (0.875,{-sqrt(3)/2})--(1.5,{-sqrt(3)/2});
\draw[thick,dashed,red] (0.5,{sqrt(3)/6})--(1,{sqrt(3)/3})--(2,{sqrt(3)/3});
\filldraw[red] (2,{sqrt(3)/3}) circle (0.1cm);
\draw[thick,directed,red] (0.5,{sqrt(3)/2})--(1,0);
\draw[thick,directed,red] (1.5,{sqrt(3)/2})--(1,0);
\draw[thick,directed,red] (1.5,{sqrt(3)/2})--(2,0);
\node at (-3.5,-2.25) {(b)};
\end{scope}

\begin{scope}[xshift=18cm]
\clip (-4,-3) rectangle (4,3);
\foreach \x in {-5,-4,...,5}{
\draw[help lines] (-5,{\x*sqrt(3)/2})--(5,{\x*sqrt(3)/2});
\draw[xshift=\x cm,help lines] (60:6)--(240:6);
\draw[xshift=\x cm,help lines] (120:6)--(300:6);
}
\filldraw[blue] (0,0) circle (0.1cm);
\draw[blue] (-0.175,-0.175) rectangle (0.175,0.175);
\filldraw[blue] (2.5,{sqrt(3)/2}) circle (0.1cm);
\draw[very thick,red] ++(0.5,{sqrt(3)/6} )--++(-60:5);
\draw[very thick,->,red] (0.875,{-sqrt(3)/2})--(1.5,{-sqrt(3)/2});
\draw[very thick,red] (0.5,{sqrt(3)/6})--(1,{sqrt(3)/3})--(2,{sqrt(3)/3});
\draw[thick,->] ++(1.5,-1.5) arc (270:460:2);
\filldraw[red] (2,{sqrt(3)/3}) circle (0.1cm);
\node at (-3.5,-2.25) {(c)};
\end{scope}

\begin{scope}[xshift=4cm,yshift=-7cm]
\clip (-4,-3) rectangle (4,3);
\foreach \x in {-5,-4,...,5}{
\draw[help lines] (-5,{\x*sqrt(3)/2})--(5,{\x*sqrt(3)/2});
\draw[xshift=\x cm,help lines] (60:6)--(240:6);
\draw[xshift=\x cm,help lines] (120:6)--(300:6);
}
\filldraw[blue] (0,0) circle (0.1cm);
\draw[blue] (-0.175,-0.175) rectangle (0.175,0.175);
\filldraw[blue] (2.5,{sqrt(3)/2}) circle (0.1cm);
\filldraw[red] (2,{sqrt(3)/3}) circle (0.1cm);
\draw[very thick,red] ++(2,{sqrt(3)/3})--++(120:3);
\draw[very thick,->,red] (1.625,{sqrt(3)})--(1,{sqrt(3)});
\draw[thick,->] ++(-2,2.5) arc (90:160:1.5);
\draw[thick,->] ++(2,-2.5) arc (270:340:1.5);
\node at (-3.5,-2.25) {(d)};
\end{scope}

\begin{scope}[xshift=13cm,yshift=-7cm]
\clip (-4,-3) rectangle (4,3);
\foreach \x in {-5,-4,...,5}{
\draw[help lines] (-5,{\x*sqrt(3)/2})--(5,{\x*sqrt(3)/2});
\draw[xshift=\x cm,help lines] (60:6)--(240:6);
\draw[xshift=\x cm,help lines] (120:6)--(300:6);
}
\filldraw[blue] (0,0) circle (0.1cm);
\filldraw[blue] (2.5,{sqrt(3)/2}) circle (0.1cm);
\draw[blue] (2.325,{sqrt(3)/2-0.175}) rectangle (2.675,{sqrt(3)/2+0.175});
\filldraw[red] (0.5,{sqrt(3)/6}) circle (0.075cm);
\draw[very thick,red] ++(0.5,{sqrt(3)/6} )--++(-60:5);
\draw[very thick,->,red] (0.875,{-sqrt(3)/2})--(1.5,{-sqrt(3)/2});
\node at (-3.5,-2.25) {(e)};
\end{scope}

\end{tikzpicture}
\caption{Computation of $G'_{(0,0),(3,1)}$ for the triangular lattice. The boxed vertex corresponds to the first argument of the derivative $G'$. The panels (a)-(e) illustrate the transformations explained in the text.}
\label{gp_ex}
\end{figure}

Let us illustrate how these properties can be used to obtain an explicit expression for the Green function derivative \cite{KW15}. Consider for instance the vertices $u_1=(0,0)$ and $u_2=(3,1)$ together with the zipper represented in the first panel of Fig.~\ref{gp_ex}. We begin by moving the head of the zipper while keeping $u_1,u_2$ fixed, that is, we add the following edges to the sum in Eq.~\eqref{tri_gp_exp}: $(1,1){-}(1,0)$, $(2,1){-}(1,0)$, $(2,1){-}(2,0)$. Then we deform and rotate the zipper counterclockwise so that it points upwards (see panels (c)-(d) in Fig.~\ref{gp_ex}). Since in doing so we drag the zipper across $u_2$, the derivative picks up an additional term $+G_{u_1,u_2}$. Finally, we rotate the whole lattice by $180^\circ$ to the left, with both $u_1,u_2$ and the zipper included, and we shift it so that $u_2$ now lies at the original position of $u_1$ (and vice versa). The antisymmetry of $G'_{u_1,u_2}$ allows one to recast $G'_{(0,0),(3,1)}$ as the sum of a finite number of terms. One finds
\begin{align}
G'_{(0,0),(3,1)}&=\frac{1}{2}\Big\{G_{(0,0),(1,1)}G_{(1,0),(3,1)}-G_{(0,0),(1,0)}G_{(1,1),(3,1)}+G_{(0,0),(2,1)}G_{(1,0),(3,1)}\nonumber \\
&\qquad-G_{(0,0),(1,0)}G_{(2,1),(3,1)}+G_{(0,0),(2,1)}G_{(2,0),(3,1)}-G_{(0,0),(2,0)}G_{(2,1),(3,1)}-G_{(0,0),(3,1)}\Big\}\nonumber \\
&=\frac{1}{2}\Big\{G(1,1)G(2,1)-G(1,0)G(2,0)+G(2,1)G(2,1)-G(1,0)G(1,0)\nonumber \\
&\qquad+G(2,1)G(1,1)-G(2,0)G(1,0)-G(3,1)\Big\}\nonumber \\
&=G_{0,0}\Big(\frac{5}{3}-\frac{4\sqrt{3}}{\pi}\Big)-\frac{107}{72}+\frac{8}{\sqrt{3}\pi}+\frac{3}{2\pi^2},
\end{align}
where the values of $G(x,y)$ given in Table \ref{tri_gf_table} have been used  together with the symmetry relation $G(3,1)=G(-2,1)$. Proceeding this way, one finds the values of the Green function derivative evaluated at the origin and its six neighbors with respect to this particular zipper, as tabulated in Table \ref{tri_gp_table}.

\begin{table}[htbp]
\centering
\large
\renewcommand{\arraystretch}{1.8}
\tabcolsep6pt
{\fontsize{0.25cm}{0.3cm}\selectfont
\begin{tabular}{|c|cccc|}
\hline
$G'_{u_1,u_2}$ & $(0,0)$ & $(1,0)$ & $(1,1)$ & $(0,1)$\\
\hline
$(0,0)$ & $0$ & ${-}\frac{2}{3}G_{0,0}{+}\frac{7}{72}$ & ${-}\frac{1}{3}G_{0,0}{+}\frac{5}{72}$ & ${-}\frac{1}{6}G_{0,0}{+}\frac{1}{24}$\\
$(1,0)$ & $\frac{2}{3}G_{0,0}{-}\frac{7}{72}$ & $0$ & $\frac{1}{3}G_{0,0}{-}\frac{5}{72}$ & $\frac{1}{2}G_{0,0}{+}\frac{1}{6}{-}\frac{\sqrt{3}}{2\pi}$\\
$(1,1)$ & $\frac{1}{3}G_{0,0}{-}\frac{5}{72}$ & ${-}\frac{1}{3}G_{0,0}{+}\frac{5}{72}$ & $0$ & $\frac{1}{6}G_{0,0}{-}\frac{1}{24}$\\
$(0,1)$ & $\frac{1}{6}G_{0,0}{-}\frac{1}{24}$ & ${-}\frac{1}{2}G_{0,0}{-}\frac{1}{6}{+}\frac{\sqrt{3}}{2\pi}$ & ${-}\frac{1}{6}G_{0,0}{+}\frac{1}{24}$ & $0$\\
$({-}1,0)$ & $G_{0,0}\left({-}\frac{1}{2}{+}\frac{\sqrt{3}}{\pi}\right){-}\frac{1}{72}$ & $G_{0,0}\left({-}\frac{7}{6}{+}\frac{\sqrt{3}}{\pi}\right){+}\frac{7}{9}{-}\frac{7}{2\sqrt{3}\pi}$ & $G_{0,0}\left({-}\frac{5}{6}{+}\frac{\sqrt{3}}{\pi}\right){-}\frac{1}{9}{+}\frac{1}{\sqrt{3}\pi}$ & $G_{0,0}\left({-}\frac{2}{3}{+}\frac{\sqrt{3}}{\pi}\right){+}\frac{1}{8}{-}\frac{1}{2\sqrt{3}\pi}$\\
$({-}1,{-}1)$ & $G_{0,0}\left(\frac{1}{2}{-}\frac{\sqrt{3}}{\pi}\right){+}\frac{1}{72}$ & $G_{0,0}\left({-}\frac{1}{6}{-}\frac{\sqrt{3}}{\pi}\right){-}\frac{2}{9}{+}\frac{2}{\sqrt{3}\pi}$ & $G_{0,0}\left(\frac{1}{6}{-}\frac{\sqrt{3}}{\pi}\right){+}\frac{5}{9}{-}\frac{5}{2\sqrt{3}\pi}$ & $G_{0,0}\left(\frac{1}{3}{-}\frac{\sqrt{3}}{\pi}\right){-}\frac{11}{36}{+}\frac{2}{\sqrt{3}\pi}$\\
$(0,{-}1)$ & ${-}\frac{1}{6}G_{0,0}{+}\frac{1}{24}$ & ${-}\frac{5}{6}G_{0,0}{+}\frac{1}{8}$ & ${-}\frac{1}{2}G_{0,0}{-}\frac{1}{6}{+}\frac{\sqrt{3}}{2\pi}$ & ${-}\frac{1}{3}G_{0,0}{+}\frac{23}{36}{-}\frac{\sqrt{3}}{\pi}$\\
\hline
\end{tabular}
\vspace{0.5cm}\\
\begin{tabular}{|c|ccc|}
\hline
$G'_{u_1,u_2}$ & $({-}1,0)$ & $({-}1,{-}1)$ & $(0,{-}1)$\\
\hline
$(0,0)$ & $G_{0,0}\left(\frac{1}{2}{-}\frac{\sqrt{3}}{\pi}\right){+}\frac{1}{72}$ & $G_{0,0}\left({-}\frac{1}{2}{+}\frac{\sqrt{3}}{\pi}\right){-}\frac{1}{72}$ & $\frac{1}{6}G_{0,0}{-}\frac{1}{24}$\\
$(1,0)$ & $G_{0,0}\left(\frac{7}{6}{-}\frac{\sqrt{3}}{\pi}\right){-}\frac{7}{9}{+}\frac{7}{2\sqrt{3}\pi}$ & $G_{0,0}\left(\frac{1}{6}{+}\frac{\sqrt{3}}{\pi}\right){+}\frac{2}{9}{-}\frac{2}{\sqrt{3}\pi}$ & $\frac{5}{6}G_{0,0}{-}\frac{1}{8}$\\
$(1,1)$ & $G_{0,0}\left(\frac{5}{6}{-}\frac{\sqrt{3}}{\pi}\right){+}\frac{1}{9}{-}\frac{1}{\sqrt{3}\pi}$ & $G_{0,0}\left({-}\frac{1}{6}{+}\frac{\sqrt{3}}{\pi}\right){-}\frac{5}{9}{+}\frac{5}{2\sqrt{3}\pi}$ & $\frac{1}{2}G_{0,0}{+}\frac{1}{6}{-}\frac{\sqrt{3}}{2\pi}$\\
$(0,1)$ & $G_{0,0}\left(\frac{2}{3}{-}\frac{\sqrt{3}}{\pi}\right){-}\frac{1}{8}{+}\frac{1}{2\sqrt{3}\pi}$ & $G_{0,0}\left({-}\frac{1}{3}{+}\frac{\sqrt{3}}{\pi}\right){+}\frac{11}{36}{-}\frac{2}{\sqrt{3}\pi}$ & $\frac{1}{3}G_{0,0}{-}\frac{23}{36}{+}\frac{\sqrt{3}}{\pi}$\\
$({-}1,0)$ & $0$ & $G_{0,0}\left({-}1{+}\frac{2\sqrt{3}}{\pi}\right){+}\frac{11}{72}{-}\frac{1}{\sqrt{3}\pi}$ & $G_{0,0}\left({-}\frac{1}{3}{+}\frac{\sqrt{3}}{\pi}\right){+}\frac{11}{36}{-}\frac{2}{\sqrt{3}\pi}$\\
$({-}1,{-}1)$ & $G_{0,0}\left(1{-}\frac{2\sqrt{3}}{\pi}\right){-}\frac{11}{72}{+}\frac{1}{\sqrt{3}\pi}$ & $0$ & $G_{0,0}\left(\frac{2}{3}{-}\frac{\sqrt{3}}{\pi}\right){-}\frac{1}{8}{+}\frac{1}{2\sqrt{3}\pi}$\\
$(0,{-}1)$ & $G_{0,0}\left(\frac{1}{3}{-}\frac{\sqrt{3}}{\pi}\right){-}\frac{11}{36}{+}\frac{2}{\sqrt{3}\pi}$ & $G_{0,0}\left({-}\frac{2}{3}{+}\frac{\sqrt{3}}{\pi}\right){+}\frac{1}{8}{-}\frac{1}{2\sqrt{3}\pi}$ & $0$\\
\hline
\end{tabular}
}
\caption{Values of the Green function derivative of the triangular lattice around the origin, with respect to the zipper shown in panel (a) of Fig.~\ref{gp_ex}.}
\label{tri_gp_table}
\end{table}


\subsection{Green function derivative for large distances}
\label{app_sec3}
On the upper half-lattice, the image method allows one to write $G^{\prime\,\mathrm{op}}$ in terms of $G'$ on the full triangular lattice, with respect to the zippers depicted in Figs.~\ref{tri_zip} and \ref{tri_uhp_cuts} for the full lattice and the half-lattice respectively. In particular, for vertices $u_i$ of the form $(a_i,p{+}b_i)$ with $a_i,b_i=o(1)$, Eq.~\eqref{tri_gp_op} yields
\begin{equation}
\begin{split}
G^{\prime\,\mathrm{op}}_{(a_1,p+b_1),(a_2,p+b_2)}&=G'_{(-a_1,-b_1),(-a_2,-b_2)}-G'_{(-a_1,-b_1),(p-a_2+b_2,2p+b_2)}\\
&\quad-G'_{(p-a_1+b_1,2p+b_1),(-a_2,-b_2)}+G'_{(p-a_1+b_1,2p+b_1),(p-a_2+b_2,2p+b_2)}.
\end{split}
\label{tri_gp_op_asy}
\end{equation}
In Subsection \ref{app_sec2}, we have shown how to compute the first term on the right-hand side, using the transformation properties of the Green function derivative $G'$. The remaining terms can in principle be evaluated in the same fashion. However, it would require the addition of $\mathcal{O}(p)$ extra edges to move the head of the zipper, and then evaluating the resulting sum for large $p$. Instead we found it more convenient to start from \eqref{gp_def} and the integral representation \eqref{tri_gfs} of the Green function. The calculation essentially follows the same steps as the corresponding one on the square lattice, made explicit in \cite{PR17}. We therefore skip the details and give the final result in a case that is relevant to \eqref{tri_gp_op_asy}, namely
\begin{equation}
\begin{split}
G'_{(0,0),(p,2p+1)}&=G_{0,0}\left(-\frac{1}{3}+\frac{1}{2\sqrt{3}\pi p}-\frac{1}{6\sqrt{3}\pi p^2}\right)+\left(\log p+\gamma+\log 6\right)\\
&\quad\times\left(\frac{1}{6\sqrt{3}\pi}-\frac{1}{24\pi^2 p}+\frac{1}{48\pi^2 p^2}\right)+\frac{1}{12\sqrt{3}\pi p}-\frac{9+2\sqrt{3}\pi}{432\pi^2 p^2}+\mathcal{O}(p^{-3}\log p).
\end{split}
\end{equation}

This expansion allows one to compute the asymptotic expansion for the second and third terms of Eq.~\eqref{tri_gp_op_asy}, that is, for $G'_{u_1,u_2}$ when only one of its arguments is close to the head of the zipper. The fourth term however corresponds to a derivative $G'_{u_1,u_2}$ where both $u_i=(p{+}a_i,2p{+}b_i)$ ($i=1,2$) are far away from the zipper. In that case, the arguments of the Green functions appearing on the right-hand side of Eq.~\eqref{tri_gp_exp} are large. Hence, we can use the asymptotic expansion \eqref{tri_gf_asy} before summing over $k$ to get the power expansion of $G'_{u_1,u_2}$ through the Euler-Maclaurin formula. For example, for $a_1=b_1=0$, $a_2=b_2=-1$, the Green function derivative reads:
\begin{equation}
\begin{split}
G'_{(p,2p),(p-1,2p-1)}&=G_{0,0}\left(\frac{1}{4\sqrt{3}\pi p}+\frac{7}{24\sqrt{3}\pi p^2}\right)+(\log p+\gamma+\log 6)\left(-\frac{1}{24\pi^2 p}-\frac{7}{144\pi^2 p^2}\right)\\
&\quad-\frac{1}{24\pi^2 p}-\frac{1}{32\pi^2 p^2}+\mathcal{O}(p^{-3}\log p).
\end{split}
\end{equation}


\subsection{Modified graphs}
Spanning tree calculations presented in Sections \ref{sec4} and \ref{sec5} require that the nodes (excluding the sink) be located around a single face. This can be achieved by defining a modified graph $\xbar{\G}$ obtained by cutting multiple (nonzipper) edges between the reference site $i$ and its neighbors, as in Figs.~\ref{tri_zip} and \ref{tri_uhp_cuts}. The line bundle Laplacian $\xbar{\bm{\Delta}}(z)$ on $\xbar{\G}$ is therefore a local symmetric modification of the Laplacian $\bm{\Delta}(z)$ on the original graph $\G$, which may be written as $\xbar{\bm{\Delta}}=\bm{\Delta}-U^{\text{t}}U$. Here the perturbation $U^{\text{t}}U$ has finite rank $r$, so $U$ can be chosen as a $r\times|\G|$ matrix. The Woodbury formula allows one to write $\xbar{\Gr}=\xbar{\bm{\Delta}}^{-1}$ and $\det\xbar{\bm{\Delta}}$ in terms of $\Gr=\bm{\Delta}^{-1}$ and $\det\bm{\Delta}$,
\begin{align}
\xbar{\Gr}&=\Gr+\Gr U^{\text{t}}\left(\mathbb{I}-U\Gr U^{\text{t}}\right)^{-1}U\Gr,\\
\det\xbar{\bm{\Delta}}&=\det\bm{\Delta}\times\det\left(\mathbb{I}-U\Gr U^{\text{t}}\right).
\end{align}
In particular, if $\xbar{\G}$ is obtained by cutting a single nonzipper edge $\{s,t\}$, the Woodbury formula reduces to the Sherman-Morrison formula. The modified Green function and its derivative then read
\begin{align}
\xbar{G}_{u,v}&=G_{u,v}-\frac{(G_{u,s}-G_{u,t})(G_{s,v}-G_{t,v})}{G_{s,s}+G_{t,t}-G_{s,t}-G_{t,s}-1},\\
\xbar{G}^{\,\prime}_{u,v}&=G'_{u,v}-\frac{(G'_{u,s}-G'_{u,t})(G_{s,v}-G_{t,v})+(G_{u,s}-G_{u,t})(G'_{s,v}-G'_{t,v})}{G_{s,s}+G_{t,t}-G_{s,t}-G_{t,s}-1}.
\end{align}
Both formulas can be used iteratively in case several edges are removed from the original graph, as an alternative to the more general Woodbury formula.



\end{document}